\documentclass[12pt]{article}

\usepackage{graphicx}

\usepackage{amsmath, amssymb}
  
\begin{document}
  
\begin{titlepage}

\def\thefootnote{\fnsymbol{footnote}}

\begin{center}

\hfill TU-840 \\
\hfill UT-HET-026 \\
\hfill March, 2009

\vspace{0.5cm}
{\Large\bf High Energy Cosmic Rays \\
from Decaying Supersymmetric Dark Matter}

\vspace{1cm}
{\large Koji Ishiwata}$^{\it (a)}$, 
{\large Shigeki Matsumoto}$^{\it (b)}$, 
{\large Takeo Moroi}$^{\it (a)}$

\vspace{1cm}

{\it $^{(a)}${Department of Physics, Tohoku University,
    Sendai 980-8578, Japan}}

\vspace{0.5cm}

{\it $^{(b)}${Department of Physics, University of Toyama, 
    Toyama 930-8555, Japan}}

\vspace{1cm}
\abstract{ 

  Motivated by the recent PAMELA and ATIC results, we calculate the
  electron and positron fluxes from the decay of
  lightest-superparticle (LSP) dark matter.  We assume that the LSP is
  the dominant component of dark matter, and consider the case that
  the $R$-parity is very weakly violated so that the lifetime of the
  LSP becomes of the order of $10^{26}\ {\rm sec}$.  We will see that,
  with such a choice of the lifetime, the cosmic-ray electron and
  positron from the decay can be the source of the anomalous $e^\pm$
  fluxes observed by PAMELA and ATIC.  We consider the possibilities
  that the LSP is the gravitino, the lightest neutralino, and scalar
  neutrino, and discuss how the resultant fluxes depend on the
  dark-matter model.  We also discuss the fluxes of $\gamma$-ray and
  anti-proton, and show that those fluxes can be consistent with the
  observed value in the parameter region where the PAMELA and ATIC
  anomalies are explained.

 }

\end{center}
\end{titlepage}

\renewcommand{\theequation}{\thesection.\arabic{equation}}
\renewcommand{\thepage}{\arabic{page}}
\setcounter{page}{1}
\renewcommand{\thefootnote}{\#\arabic{footnote}}
\setcounter{footnote}{0}

\section{Introduction}
\label{sec:intro}
\setcounter{equation}{0}

In astrophysics, the existence of dark matter is almost conclusive.
According to the recent survey of WMAP \cite{Hinshaw:2008kr}, it
accounts for 23 \% of the total energy density in the universe.  In
the standard model of particle physics, however, there does not exist
candidate for dark matter, which is one of the reasons to call for new
physics beyond the standard model.  Supersymmetry (SUSY) is a
promising model which can give an answer to the question; in the
framework of SUSY, lightest superparticle (LSP) is a viable candidate
for dark matter.

The fluxes of high energy cosmic rays give information about the
properties of dark matter.  In the recent years, accuracy of the
measurements of the fluxes have been significantly improved.  In
particular, anomalous signals are reported by PAMELA
\cite{Adriani:2008zr} and ATIC \cite{:2008zzr} in the observations of
high energy cosmic-ray positron and electron.  The PAMELA and ATIC
results have attracted many attentions because the anomalies may
indicate an unconventional nature of dark matter.  In fact, a sizable
number of dark-matter models are proposed to explain the anomalies
after the announcements of the PAMELA and ATIC results.  Roughly
speaking, the possibilities to produce such high energy electron and
positron are categorized into two: the decay and the annihilation of
dark matter.\footnote
{The other possibilities to enhance the $e^\pm$ fluxes considered
 using nearby pulsars \cite{Hooper:2008kg}.}
(For early attempts to calulate the spectra of cosmic-ray $e^{\pm}$,
see
\cite{Nomura:2008ru,Yin:2008bs,Ishiwata:2008cv,Bai:2008jt,Chen:2008md,Hamaguchi:2008rv,Ponton:2008zv,Ibarra:2008jk,Chen:2008qs,Arvanitaki:2008hq,Hamaguchi:2008ta,Gogoladze:2009kv,Hamaguchi:2009sz,Nardi:2008ix,Huh:2008vj,Chen:2008dh,Chen:2009mf}
and
\cite{Cirelli:2008pk,Cholis:2008qq,Feldman:2008xs,Ishiwata:2008cv,Fox:2008kb,Bergstrom:2008gr,Barger:2008su,Nelson:2008hj,Harnik:2008uu,Hisano:2009rc} 
for decaying and annihilating dark matter, respectively.)  In general,
the latter case has the difficulty to reproduce the anomalous
cosmic-ray positron and electron fluxes without large enhancement
factor, called boost factor.\footnote
{However, cosmic-ray $e^\pm$ fluxes may be enhanced without large
  boost factor with the Breit-Wigner enhancement of the annihilation
  cross section \cite{Ibe:2008ye}, Sommerfelt enhancement
  \cite{Hisano:2003ec}, or with a nearby clump of dark matter
  \cite{Hooper:2008kv}.}
In the former case, on the other hand, the observed anomalies can be
well explained with the appropriate choice of the lifetime of dark
matter especially in leptonically decaying scenarios.

In usual supersymmetric scenario, $R$-parity conservation is assumed,
which protects LSP from decaying into standard model particles and
makes it a viable candidate for dark matter.  If we consider the case
that $R$-parity is violated, LSP is no loner stable; however, if
$R$-parity violation (RPV) is weak enough, the lifetime of the LSP can
be much longer than the present age of the universe and LSP can play
the role of dark matter \cite{Takayama:2000uz}.  In addition, when the
size of the RPV is properly chosen to give the lifetime of $O(10^{26}\
{\rm sec})$, produced cosmic-ray positron gives excellent agreement
with PAMELA data \cite{Ishiwata:2008cv}.

In this paper, we calculate fluxes of cosmic-ray positron, electron,
$\gamma$-ray, and anti-proton in various LSP dark matter scenarios,
paying particular attentions to the results given by PAMELA and ATIC.
We consider the cases where the LSP dark matter is unstable, assuming
that $R$-parity is (very weakly) violated.  Then, we compare the
calculated fluxes with the results of observations.  We will see that
the PAMELA and ATIC anomalies are simultaneously explained if the
lifetime of the LSP dark matter is $O(10^{26}\ {\rm sec})$ and the
mass is $\sim 1 - 1.5\ {\rm TeV}$.  In addition, in some cases,
$\gamma$-ray and anti-proton are also produced by the decay of the
LSP.  We will see that, taking account of the uncertainties in the
Galaxy and propagation models as well as the error in the
observations, the scenarios are not excluded by the observations of
$\gamma$-ray and anti-proton fluxes yet.

The organization of this paper is as follows.  In the next Section,
we explain the basic procedures to calculate the high energy cosmic-ray
fluxes.  Then, we discuss the cosmic-ray fluxes for the cases where
the LSP is the gravitino, the lightest neutralino, and the sneutrino
in Sections \ref{sec:gravitino}, \ref{sec:neutralino}, and
\ref{sec:sneutrino}, respectively.  Section \ref{sec:conclusions} is
devoted for conclusions and discussion.

\section{Basic Setup}
\label{sec:setup}
\setcounter{equation}{0}

As we have discussed in introduction, we consider the case where the
LSP is unstable and has lifetime much longer than the present age of
the universe so that most of the LSPs produced in the early universe
survive until today.  Then, if the relic density of the LSP is right
amount, the LSP can be dark matter. In such a case, the LSP dark
matter becomes the source of high energy cosmic rays.

The fluxes of the cosmic rays from the decay of the relic LSP depends
on what the LSP is, and how the LSP decays. The LSP should be an
electrically neutral particle and we consider the following three
important cases:
\begin{itemize}
\item Gravitino LSP
\item Neutralino LSP 
\item Sneutrino LSP
\end{itemize}
In addition, we adopt $R$-parity violation so that the LSP becomes
unstable.  Then, the lifetime of the LSP becomes longer as the
coupling constants for the RPV interactions become smaller. We
therefore treat the lifetime of the LSP as a free parameter in the
following analysis. To be specific, we assume that the $R$-parity
violation is so weak that the lifetime of the LSP becomes much longer
than the present cosmic time.

\subsection{Sources of the Cosmic Rays}

With the decay of relic LSP, energetic particles (in particular,
$\gamma$, $e^\pm$, and anti-proton $\bar{p}$) are produced. The
production rate of the energetic particle $I$ is given by
\begin{eqnarray}
  Q_I (E, \vec{x}) = 
  \frac{1}{\tau_{\rm LSP}} n_{\rm LSP}(\vec{x})
  \left[ \frac{d N_I}{d E} 
  \right]_{\rm dec},
  \label{source_I}
\end{eqnarray}
where $\left[dN_I/dE\right]_{\rm dec}$ is the energy distribution of
particle $I$ from the single LSP decay process. We have used the
PYTHIA package \cite{Sjostrand:2006za} to calculate
$\left[dN_I/dE\right]_{\rm dec}$. In addition, $\tau_{\rm LSP}$ and
$n_{\rm LSP}$ are the lifetime and number density of the LSP at the
present universe, respectively. In the calculation of the cosmic-ray
spectrum, the origin of the dark-matter LSP is unimportant.  Even
though it is often assumed that the relic density of the LSP is
thermally determined, non-thermal production of the LSP dark matter is
also possible \cite{Moroi:1999zb}.  We thus do not specify the origin
of the relic LSP in the following analysis, and set $n_{\rm LSP}$ to
be $\Omega_{\rm LSP} = \Omega_{\rm DM}$, where $\Omega_{\rm LSP}$ and
$\Omega_{\rm DM}$ are the density parameters of the LSP and dark
matter, respectively.\footnote
{Even if $\Omega_{\rm LSP}<\Omega_{\rm DM}$, the relic LSP can still
  be the source of high energy cosmic rays. Then, the fluxes can be
  obtained by rescaling the lifetime as $\tau_{\rm LSP} \rightarrow
  (\Omega_{\rm LSP}/\Omega_{\rm DM})\tau_{\rm LSP}$.}
Then, the mass density of dark matter is given by $\rho_{\rm
  DM}(\vec{x})=m_{\rm LSP}n_{\rm LSP}(\vec{x})$. In calculating the
  fluxes of high-energy cosmic rays, we adopt the Navarro-Frank-White
  (NFW) mass density profile \cite{Navarro:1996gj}:
\begin{eqnarray}
  \rho_{\rm NFW} (\vec{x}) = \rho_\odot 
  \frac{r_\odot (r_c + r_\odot)^2}{r (r_c + r)^2},
  \label{eq:nfw}
\end{eqnarray}
where $\rho_\odot\simeq 0.30\ {\rm GeV/cm^3}$ is the local halo
density around the solar system, $r_c \simeq 20\ {\rm kpc}$ is the
core radius of the dark matter profile, $r_\odot \simeq 8.5\ {\rm
  kpc}$ is the distance between the Galactic center and the solar
system, and $r$ is the distance from the Galactic center. Using the
$Q_I$ given in Eq.\ \eqref{source_I} as a source term, we solve the
propagation equations for individual particles. The diffusion zone is
approximated as a cylinder with half-height $L$ and radius $R=20\ {\rm
  kpc}$.
\subsection{Electron \& Positron fluxes}

In order to calculate the fluxes of electron and positron from the LSP
decay, we derive a static solution of the following diffusion
equation:
\begin{eqnarray}
  \frac{\partial  f_{e^\pm}(E,\vec{x})}{\partial t}
  = K_{e^\pm} (E) \nabla^2 f_{e^\pm}(E,\vec{x})
  + \frac{\partial}{\partial E}\left[ b(E) 
    f_{e^\pm}(E,\vec{x}) \right]
  + Q_{e^\pm}(E,\vec{x}),
  \label{DiffEq_epm}
\end{eqnarray}
with the condition $f_{e^\pm}=0$ at the boundary of the diffusion
zone, where $f_{e^\pm}$ is the number density of $e^\pm$ per unit
energy.  Our basic procedure to solve the diffusion equation
\eqref{DiffEq_epm} is explained in Appendix \ref{appendix:greenFn}.

We approximate the function $K_{e^\pm}$ as \cite{Delahaye:2007fr}
\begin{eqnarray}
  K_{e^\pm}=K_{e^\pm}^{(0)} E_{\rm GeV}^\delta,
\end{eqnarray}
with $E_{\rm GeV}$ being energy in units of GeV, while the energy-loss
rate $b$ is given by
\begin{eqnarray}
  b=1.0\times 10^{-16} E_{\rm GeV}^2\ {\rm GeV/sec}.
\end{eqnarray}
We adopt three sets of diffusion parameters, which are summarized in
Table \ref{table:propagation_e}. The MED set gives the best-fit value
in the boron-to-carbon ratio (B/C) analysis, while the maximal and
minimal positron fractions are expected in the M1 and M2 sets without
conflicting the B/C analysis.

\begin{table}[t]
  \begin{center}
    \begin{tabular}{lccc}
      \hline \hline
      & M1 & MED & M2
      \\
      \hline
      $L\ {\rm [kpc]}$  & 15   & 4   &  1\\
      $\delta$ & 0.46  &  0.70 & 0.55\\
      $K_{e^\pm}^{(0)}\ {\rm [kpc^2/Myr]}$
      & 0.0765 &  0.0112 & 0.00595\\
      \hline \hline
    \end{tabular}
    \caption{\small Parameter sets for the propagation model of $e^\pm$.}
    \label{table:propagation_e}
  \end{center}
\end{table}

Once $f_{e^\pm}$ are given by solving the above equation, the fluxes
can be obtained as
\begin{eqnarray}
  \left[ \Phi_{e^\pm} (E) \right]_{\rm DM} =
  \frac{c}{4 \pi} f_{e^\pm}(E, \vec{x}_{\odot}), 
\end{eqnarray}
where $\vec{x}_{\odot}$ is the location of the solar system, and $c$
is the speed of light. In order to calculate the total fluxes of
$e^\pm$, we also have to estimate the background fluxes. In our study,
we adopt the following fluxes for cosmic-ray electrons and positrons
produced by collisions between primary protons and interstellar medium
in our galaxy \cite{Baltz:1998xv}:
\begin{eqnarray}
  \left[ \Phi_{e^-} \right]_{\rm prim}
  &=& 
  \frac{0.16 E_{\rm GeV}^{-1.1}}
  {1+11E_{\rm GeV}^{0.9}+3.2E_{\rm GeV}^{2.15}}
  ({\rm GeV \ cm}^2 \  {\rm sec \  str})^{-1},
  \\
  \left[ \Phi_{e^-} \right]_{\rm sec}
  &=& 
  \frac{0.70E_{\rm GeV}^{0.7}}
  {1+110E_{\rm GeV}^{1.5}+600E_{\rm GeV}^{2.9}+580E_{\rm GeV}^{4.2}}
  ({\rm GeV \ cm}^2 \ {\rm sec \  str})^{-1},
  \\
  \left[ \Phi_{e^+} \right]_{\rm sec}
  &=& \frac{4.5E_{\rm GeV}^{0.7}}
  {1+650E_{\rm GeV}^{2.3}+1500E_{\rm GeV}^{4.2}}
  ({\rm GeV \ cm}^2 \ {\rm sec \  str})^{-1}.
  \label{BG_e+}
\end{eqnarray}
With these backgrounds, the total fluxes are obtained as
\begin{eqnarray}
  \left[ \Phi_{e^+} \right]_{\rm tot}
  &=& \left[ \Phi_{e^+} \right]_{\rm DM}
  + \left[ \Phi_{e^+} \right]_{\rm sec},
  \label{Phi_e+}
  \\
  \left[ \Phi_{e^-} \right]_{\rm tot}
  &=& 
  \left[ \Phi_{e^-} \right]_{\rm DM}
  + \left[ \Phi_{e^-} \right]_{\rm prim}
  + \left[ \Phi_{e^-} \right]_{\rm sec}.
  \label{Phi_e-}
\end{eqnarray}
Using the fluxes defined above, the positron fraction, which is
measured by the PAMELA, is defined as
\begin{eqnarray}
  R_{e^+} = \frac
  {[ \Phi_{e^+} (E) ]_{\rm tot}}
  {[ \Phi_{e^-} (E) ]_{\rm tot} + [ \Phi_{e^+} (E) ]_{\rm tot}}.
\end{eqnarray}

\subsection{Anti-proton flux}

The flux of $\bar{p}$ from the LSP decay is obtained by solving the
diffusion equation:
\begin{eqnarray}
  \frac{\partial f_{\bar{p}}(E,\vec{x})}{\partial t}
  &=& 
  K_{\bar{p}} (E) \nabla^2 f_{\bar{p}} (E,\vec{x})
  - \frac{\partial}{\partial z}
  \left[ V_c {\rm sign} (z) f_{\bar{p}} (E,\vec{x}) \right]
  \nonumber \\ &&
  - 2 h \delta(z) \Gamma_{\rm ann} f_{\bar{p}} (E,\vec{x})
  + Q_{\bar{p}}(E,\vec{x}),
\end{eqnarray}
where $z$ is the distance from the Galactic plane. Here, $V_c$ is the
convection velocity, $h$ is the half height of the thin Galactic disc,
which is taken to be $h=100\ {\rm pc}$, and $\Gamma_{\rm ann}$ is the
annihilation rate of $\bar{p}$ in the Galactic disc, which is given by
\begin{eqnarray}
  \Gamma_{\rm ann} = (n_{\rm H} + 4^{2/3} n_{\rm He}) 
  \sigma_{p\bar{p}} v_{\bar{p}},
\end{eqnarray}
where we use the number density of Hydrogen and Helium in the Galactic
disc to be $n_{\rm H}= 1\ {\rm cm}^{-3}$ and $n_{\rm He}= 0.07n_{\rm
  H}$. The cross section $\sigma_{p\bar{p}}$ is given by
\cite{Tan:1983de,Protheroe:1981gj}
\begin{eqnarray}
  \sigma_{p\bar{p}} = \left\{ 
    \begin{array}{ll}
      661(1 + 0.0115 T_{\rm GeV}^{-0.774} 
      - 0.948 T_{\rm GeV}^{0.0151})\ {\rm mb} &
      ~:~ T_{\rm GeV} < 14.6\ {\rm GeV} \\
      36 T_{\rm GeV}^{-0.5}\ {\rm mb} &
      ~:~ T_{\rm GeV} \geq 14.6\ {\rm GeV} 
    \end{array} \right.,
\end{eqnarray}
with $T_{\rm GeV}$ being the kinetic energy of the anti-proton in
units of GeV. In addition, as in the case of $e^\pm$, the function
$K_{\bar{p}}$ is parametrized as
\begin{eqnarray}
  K_{\bar{p}}=K_{\bar{p}}^{(0)}\beta_{\bar{p}}p_{\rm GeV}^\delta,
\end{eqnarray}
where $\beta_{\bar{p}}$ is the velocity of the anti-proton, and
$p_{\rm GeV}$ is the momentum in units of GeV. Parameter sets for the
diffusion equation used in our analysis are summarized in Table
\ref{table:propagation_pbar}. Again, the MED set gives the best-fit to
the B/C analysis, while the maximal and minimal anti-proton fluxes are
expected in the MAX and MIN sets. Once $f_{\bar{p}}$ is obtained, the
anti-proton flux at the solar system is calculated as
\begin{eqnarray}
  \left[ \Phi_{\bar{p}} (E) \right]_{\rm DM} = 
    \frac{c\beta_{\bar{p}}}{4 \pi} f_{\bar{p}} (E, \vec{x}_{\odot}).
\end{eqnarray}

\begin{table}[t]
  \begin{center}
    \begin{tabular}{lccc}
      \hline \hline
      &  MAX & MED & MIN
      \\
      \hline
      $L\ {\rm [kpc]}$  & 15   & 4   &  1\\
      $\delta$ & 0.46  &  0.70 & 0.85\\
      $K_{\bar{p}}^{(0)}\ {\rm [kpc^2/Myr]}$
      & 0.0765 &  0.0112 & 0.0016\\
      $V_{\rm c}\ {\rm [km/s]}$
      & 5 &  12 & 13.5\\
      \hline \hline
    \end{tabular}
    \caption{\small Parameter sets for the propagation models of $\bar{p}$.}
    \label{table:propagation_pbar}
  \end{center}
\end{table}

\subsection{$\gamma$-ray flux}

The flux of the $\gamma$-ray is calculated by the sum of two
contributions:
\begin{eqnarray}
  \left[\Phi_{\gamma} \right]_{\rm DM} = 
  \left[\Phi_{\gamma} \right]_{\rm cosmo} + 
  \left[\Phi_{\gamma} \right]_{\rm halo},
\end{eqnarray}
where the first and second terms in the right-hand side are fluxes of
$\gamma$-ray from cosmological distance and that from the Milky Way halo,
respectively.\footnote
{$\gamma$-ray may be also produced by the inverse Compton (IC)
    scattering process.  
    We have estimated the $\gamma$-ray flux from the IC process with
    the cosmic microwave background radiation in the sky region
    $10^{\circ}<b<20^{\circ}$ (with $b$ here being Galactic longitude)
    to compare with the preliminary FERMI results, and found that the
    flux is much smaller than the FERMI data.  Detailed study of the
    $\gamma$-ray flux from the IC process will be given elsewhere
    \cite{IshiwataMatsumotoMoroi}.}

The flux from cosmological distance $\left[\Phi_{\gamma} \right]_{\rm
cosmo}$ is obtained as
\begin{eqnarray}
 \left[\Phi_\gamma \right]_{\rm cosmo}
 =
 \frac{1}{m_{{\rm LSP}} \tau_{{\rm LSP}}} 
 \int_E^{\infty} dE^{\prime} G_{\gamma}(E,E^{\prime})  
 \left[\frac{dN_\gamma}{dE^{\prime}}\right]_{\rm dec}.
\end{eqnarray}
Here, the propagation function of $\gamma$-ray turns out to be
\begin{eqnarray}
  G_{\gamma}(E,E^{\prime})
  = \frac{c \rho_c \Omega_{\rm LSP} }{4 \pi H_0 \Omega_M^{1/2}}
  \frac{1}{E} \left( \frac{E}{E^{\prime}} \right)^{{3/2}}
  \frac {1}{\sqrt{1+ \Omega_{\Lambda}/\Omega_M (E/E^{\prime})^3}},
\end{eqnarray}
where $H_0$ is present Hubble expansion rate, $\rho_c$ is critical
density, while $\Omega_M \simeq 0.137\ h^{-2}$ and
$\Omega_{\Lambda}\simeq 0.721$ (with $h \simeq 0.701$) are density
parameters of total matter and dark energy, respectively
\cite{Hinshaw:2008kr}.  \footnote
{ We found a typo in Eq.(4.3) of
\cite{Ishiwata:2008cu}.  $\Omega_M$ in the first factor of right-hand
side (i.e., $c \rho_c \Omega_{3/2}/4 \pi H_0 \Omega_M$) should be
$\Omega_M^{1/2}$.  }

On the other hand, the flux from the Milky Way
halo $\left[\Phi_{\gamma} \right]_{\rm halo}$ is calculated as
\begin{eqnarray}
  \left[\Phi_{\gamma} \right]_{\rm halo}
  = \frac{1}{m_{{\rm LSP}} \tau_{{\rm LSP}}} 
  \frac{1}{4 \pi}
  \left[\frac{dN_\gamma}{dE^{\prime}}\right]_{\rm dec}
  \left\langle  
    \int_{\rm l.o.s.} \rho_{\rm DM}(\vec{l}) \ d \vec{l}
  \right\rangle_{\rm dir},
  \label{J(gamma)_halo}
\end{eqnarray}
where the integration should be understood to extend over the line of
sight (l.o.s.) and $\langle \cdots \rangle_{\rm dir}$ means averaging
over the direction.  In the EGRET observation \cite{Sreekumar:1997un},
the signal from the Galactic disc is excluded in order to avoid the
noise. Thus, in order to compare our numerical results with the EGRET
observation, we exclude the region within $\pm 10^{\circ}$ around the
Galactic disc in the averaging.

For the background flux against the signal, we adopt the following
flux formula which is estimated from the EGRET observation in the
energy range $0.05 \ {\rm GeV} \leq E \leq 0.15$ GeV
\cite{Ishiwata:2008cu}:
\begin{eqnarray}
  E^2 \left[\Phi_\gamma\right]_{\rm BG}
  \simeq 5.18 \times 10^{-7}\ 
  ({\rm cm}^2\ {\rm sec} \ {\rm str})^{-1} \ {\rm GeV} 
  \times \left( \frac{E}{{\rm GeV}} \right)^{-0.449}.
  \label{BG_gamma}
\end{eqnarray}
We have assumed that the spectrum of the background flux from
astrophysical origins follows a power law, and its behavior can be
extracted to the high energy region. The total $\gamma$-ray spectrum
is then given by
\begin{eqnarray}
  \left[ \Phi_\gamma \right]_{\rm tot} = 
  \left[ \Phi_\gamma \right]_{\rm DM}
  + 
  \left[ \Phi_\gamma \right]_{\rm BG}.
  \label{J_gamma}
\end{eqnarray}

\section{Gravitino LSP}
\label{sec:gravitino}
\setcounter{equation}{0}

Now, we are at the position to discuss the fluxes of $e^\pm$ for
individual dark matter scenarios.  The first example is the case where
the gravitino, which is denoted as $\psi_\mu$, is the LSP and hence is
dark matter.  The model discussed here is the same as that given in
\cite{Ishiwata:2008cu, Buchmuller:2007ui, Ibarra:2008qg}, and we
consider the following RPV interaction
\begin{eqnarray}
  {\cal L}_{\rm RPV} 
  = B_i \tilde{L}_i H_u + m^2_{\tilde{L}_i H_d} \tilde{L}_i H^*_d 
  + {\rm h.c.},
  \label{L_RPV}
\end{eqnarray}
where $\tilde{L}_i$ is left-handed slepton doublet in $i$-th
generation, while $H_u$ and $H_d$ are up- and down-type Higgs boson
doublets, respectively.  (In the present analysis, $H_d$ and
$\tilde{L}_i$ are defined in the frame in which the bi-linear
$R$-parity violating terms in the superpotential vanish.)  

The effects of the RPV with \eqref{L_RPV} is parametrized by the VEV
of the sneutrino fields $\tilde{\nu}_{i}$.  To parametrize the VEV of
$\tilde{\nu}_{i}$, we define
\begin{eqnarray}
  \kappa_i \equiv \frac{\langle \tilde{\nu}_{i} \rangle}{v}
  = \frac{B_i \sin\beta + m^2_{\tilde{L}_i H_d} \cos\beta}
  {m^2_{\tilde{\nu}_{i}}},
\end{eqnarray}
where $v\simeq 174\ {\rm GeV}$ is the VEV of standard-model-like Higgs
boson, $\tan \beta = \langle H^0_u \rangle / \langle H^0_d \rangle$,
and $m_{\tilde{\nu}_{i}}$ is the mass of $\tilde{\nu}_{i}$.  Since we
are interested in the case of very long lifetime, we consider the case
that $\kappa_i\ll 1$.

With the RPV operator given in Eq.\ \eqref{L_RPV}, the gravitino
decays as $\psi_{\mu} \rightarrow \gamma \nu_i$, $Z\nu_i$, $Wl_i$, and
$h\nu_i$.  (For detailed calculations of the decay rates for these
processes, see in \cite{Ishiwata:2008cu}.)  In particular, in the
limit that the gravitino is much heavier than the weak bosons, the
following relation holds: $\Gamma_{\psi_\mu \rightarrow Z \nu_i}\simeq
\Gamma_{\psi_\mu \rightarrow h \nu_i}\simeq\frac{1}{2}\Gamma_{\psi_\mu
\rightarrow W l_i}$, and the process $\psi_\mu \rightarrow\gamma\nu_i$
is suppressed.  Then, once the relic gravitino decays, the produced
$l_i$ (as well as the weak and Higgs bosons) becomes the source of
cosmic-ray electron and positron.  In addition, with the hadronic
decay of the weak and Higgs bosons, energetic anti-proton and
$\gamma$-ray are also produced.  Fluxes of these particles have been
measured, and in the following, we compare the expected fluxes with
the results of observations.

As we have seen, the fluxes of the cosmic rays originating from the
gravitino decay is inversely proportional to the lifetime of the
gravitino $\tau_{3/2}$.  With the RPV interaction given in Eq.\
\eqref{L_RPV}, the lifetime of the gravitino is approximately given by
\begin{eqnarray}
  \tau_{3/2}\simeq 6 \times 10^{25}\ {\rm sec} \times
  \left( \frac{\kappa}{10^{-10}} \right)^{-2} 
  \left( \frac{m_{3/2}}{1\ {\rm TeV}} \right)^{-3},
\end{eqnarray}
where
\begin{eqnarray}
  {\kappa}^2 \equiv \sum_i \kappa_i^2.
\end{eqnarray}

Now, we show the cosmic-ray fluxes originating from the gravitino
decay.  First, we consider the fluxes of $e^\pm$, motivated by their
observed anomalous fluxes recently reported by the PAMELA and ATIC
experiments.  In order to discuss the preferred lifetime to explain
the anomalous positron fraction observed by the PAMELA experiment, we
define the $\chi^2$ variable as
\begin{eqnarray}
  \chi^2 = \sum_i 
  \frac{(R_{e^+,i}^{\rm (obs)} - R_{e^+,i}^{\rm (th)})^2}
  {\delta R_{e^+,i}^{{\rm (obs)}2}},
\end{eqnarray}
where $R_{e^+,i}^{\rm (obs)}$ and $R_{e^+,i}^{\rm (th)}$ are positron
fraction in $i$-th bin measured by the PAMELA and that predicted in
the unstable dark matter scenario, respectively, and $\delta
R_{e^+,i}^{{\rm (obs)}}$ is the error in the observed fraction.  Since
the positron flux in the low-energy region is sensitive to the
background fluxes, we only use the data points with $E\geq 15\ {\rm
GeV}$ ($5$ data points) in the calculation of $\chi^2$.  (As we will
discuss, the positron fraction depends on the background.  We use the
$\chi^2$ variable just to estimate the preferred value of the lifetime
to explain the PAMELA results.)

The best-fit lifetime to explain the PAMELA anomaly depends on the
gravitino mass, the flavor of the final-state lepton, and the
propagation parameters.  In our analysis, for simplicity, we consider
the cases where the LSP dominantly decays into fermions in one of the
three generations.  Concerning the other parameters, we will discuss
how the resultant fluxes depend on them.

\begin{figure}
    \begin{center}
      \includegraphics[scale=1.1]{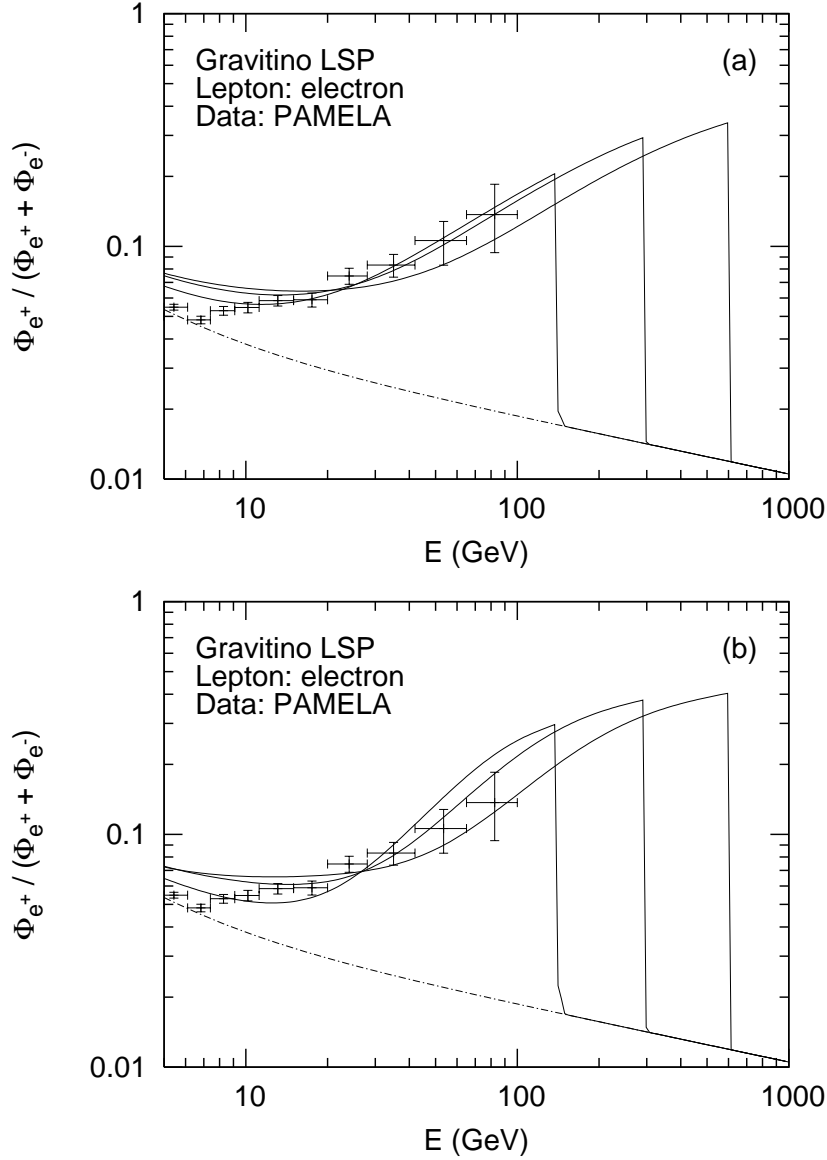}
      \caption{\small Positron fractions in (a) MED and (b) M2 models
        for the case where the gravitino dominantly decays to the
        first-generation lepton.  Dot-dashed line is the positron
        fraction calculated only by the background fluxes.  Here, we
        take $m_{3/2}=300\ {\rm GeV}$, 600 GeV, and 1.2 TeV (from left
        to right) with $2.0\ \times 10^{26}\ {\rm sec}$, $1.1\ \times
        10^{26}\ {\rm sec}$, and $8.6\ \times 10^{25}\ {\rm sec}$
        ($9.3\ \times 10^{25}\ {\rm sec}$, $5.0\ \times 10^{25}\ {\rm
        sec}$, and $4.3\ \times 10^{25}\ {\rm sec}$) in MED (M2)
        model, respectively.  }
\label{fig:positronfraction_e}
    \end{center}
\end{figure}

For the case where the gravitino decays only into the first-generation
lepton, we plot the positron fraction in Fig.\
\ref{fig:positronfraction_e} with the best-fit lifetime.  Here, we
show results with adopting the MED and M2 propagation models, because
the results with the M1 and MED models are almost the same.  For the
MED (M2) propagation model, the best-fit lifetime is given by $2.0\
\times 10^{26}\ {\rm sec}$, $1.1\ \times 10^{26}\ {\rm sec}$, and
$8.6\ \times 10^{25}\ {\rm sec}$ ($9.3\ \times 10^{25}\ {\rm sec}$,
$5.0\ \times 10^{25}\ {\rm sec}$, and $4.3\ \times 10^{25}\ {\rm
sec}$) for $m_{3/2}=300\ {\rm GeV}$, $600\ {\rm GeV}$, and $1.2\ {\rm
TeV}$, respectively.

\begin{figure}
    \begin{center}
      \includegraphics[scale=1.1]{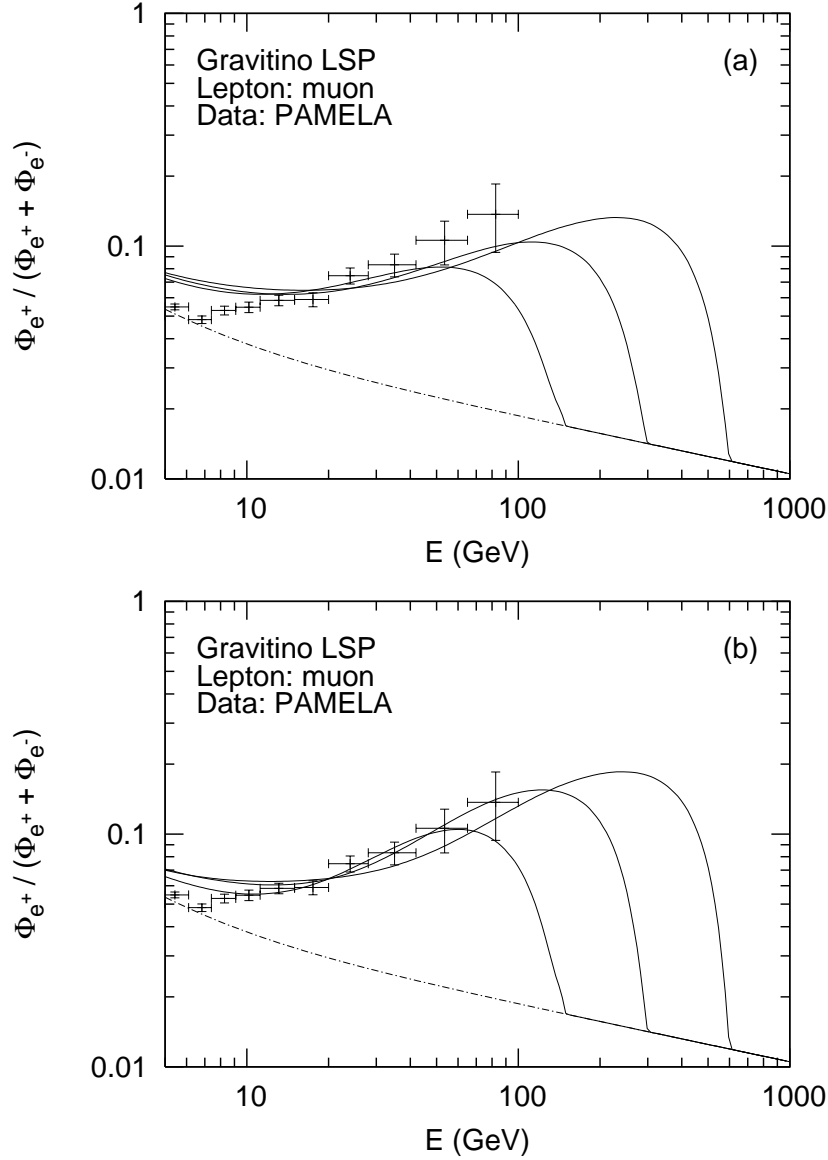}
      \caption{\small Same as Fig.\ \ref{fig:positronfraction_e}, but
        for the case where the gravitino dominantly decays to
        second-generation lepton.  Here, we take $\tau_{3/2}=1.5
        \times 10^{26}\ {\rm sec}$, $1.1 \times 10^{26}~{\rm sec}$,
        and $8.6 \times 10^{25}$ sec ($9.3\ \times 10^{26}\ {\rm
        sec}$, $5.8\ \times 10^{26}\ {\rm sec}$, and $5.0\ \times
        10^{25}\ {\rm sec}$) in MED (M2) model, which are the best-fit
        lifetime.  }
      \label{fig:positronfraction_mu}
    \end{center}
\end{figure}

\begin{figure}
    \begin{center}
      \includegraphics[scale=1.1]{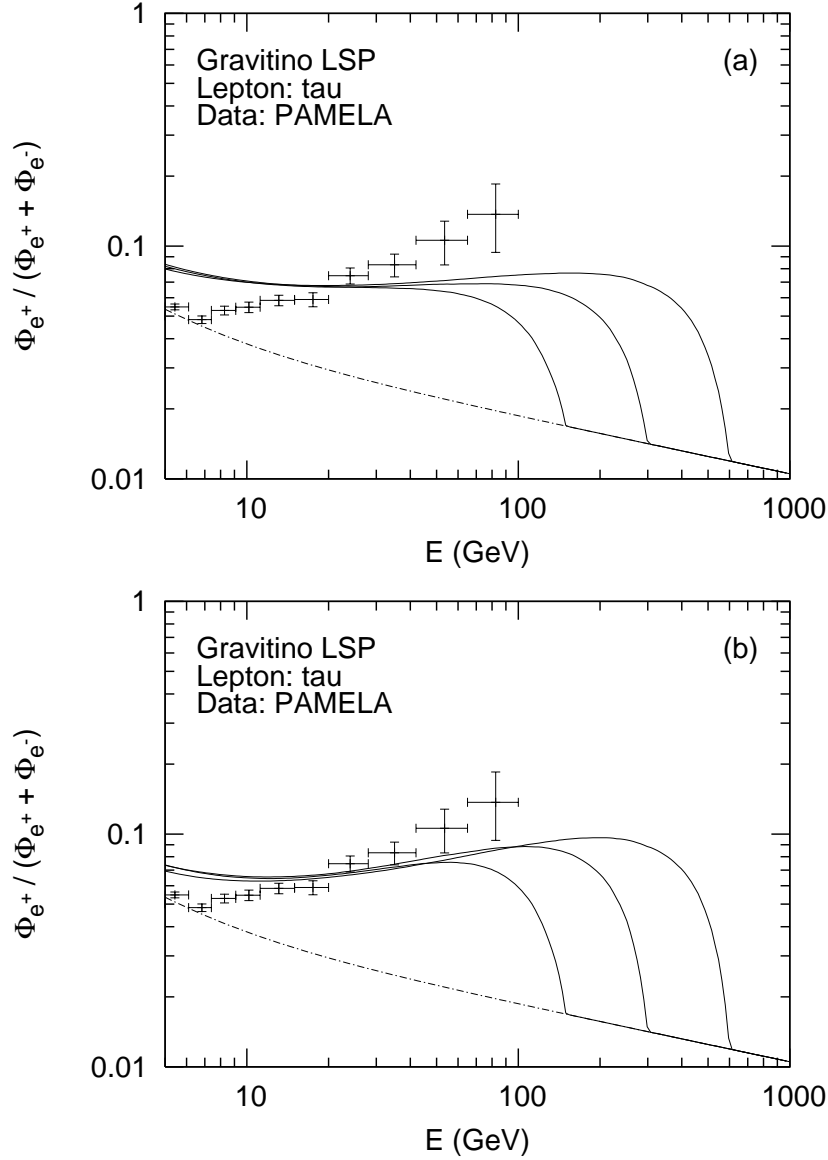}
      \caption{\small Same as Fig.\ \ref{fig:positronfraction_ee}, but
        for the case where the gravitino dominantly decays to
        third-generation lepton.  Here, we take $\tau_{3/2}=9.3 \times
        10^{25}\ {\rm sec}$, $8.6 \times 10^{25}~{\rm sec}$, and $7.9
        \times 10^{25}$ sec ($6.3\ \times 10^{25}\ {\rm sec}$, $5.4\
        \times 10^{25}\ {\rm sec}$, and $5.4\ \times 10^{25}\ {\rm
        sec}$) in MED (M2) model, which are the best-fit lifetime.  }
      \label{fig:positronfraction_tau}
    \end{center}
\end{figure}

The positron fraction for the cases where the gravitino decays only
into second- and third-generation lepton are shown in Figs.\
\ref{fig:positronfraction_mu} and \ref{fig:positronfraction_tau},
respectively.  Here, we show the results with the propagation models
MED and M2. We can see that the predicted positron fraction well
agrees with the PAMELA result if the lifetime is properly chosen.  In
particular, when the final-state lepton is first- (second-)
generation, the fit is excellent irrespective of the gravitino mass
with MED (M2) propagation model.  We note here that, for $e^-$, the
background flux is significantly larger than the signal flux, while
the dark matter contribution dominates for $e^+$.  Thus, the result is
sensitive to the choice of background because we plot the positron
fraction.  However, the theoretical calculation of the positron
fraction contains parameters both in the particle-physics model (i.e.,
the lifetime and the flavor of the final-state leptons) and in the
propagation model.  Thus, we believe that positron fraction observed
by the PAMELA can be explained in the present scenario with other
choice of the background fluxes.  For example, even if the
normalization of the background $e^-$ flux is changed, we can obtain
almost the same positron fraction by varying the lifetime.

\begin{figure}
    \begin{center}
      \includegraphics[scale=1.1]{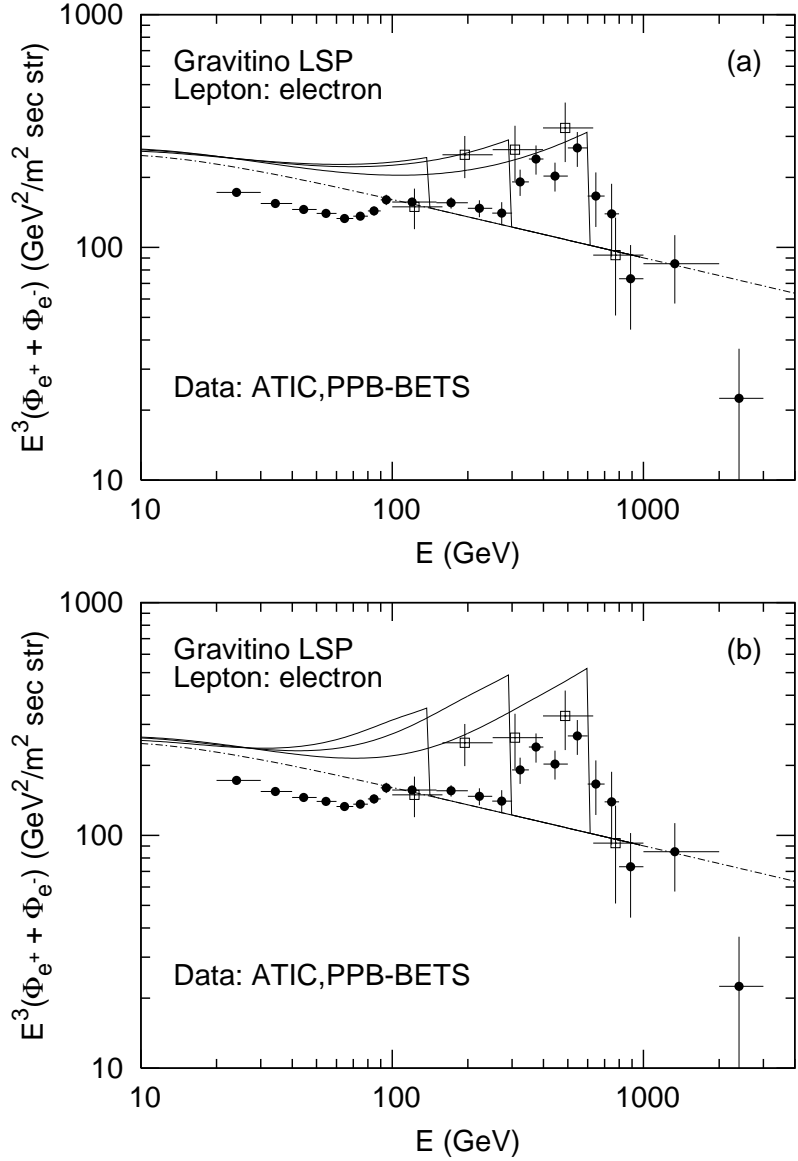}
      \caption{\small Total fluxes of positron and electron in (a) MED
        and (b) M2 models for the case where the gravitino dominantly
        decays to the first-generation lepton.  Dot-dashed line is the
        background flux.  Here, we take the the same mass and lifetime
        as Fig.\ \ref{fig:positronfraction_e}, and also plot PPB-BETS
        data \cite{Torii:2008xu}.  }
        \label{fig:positronflux_e}
    \end{center}
    \vspace{-0.5cm}
\end{figure}

\begin{figure}
    \begin{center}
      \includegraphics[scale=1.1]{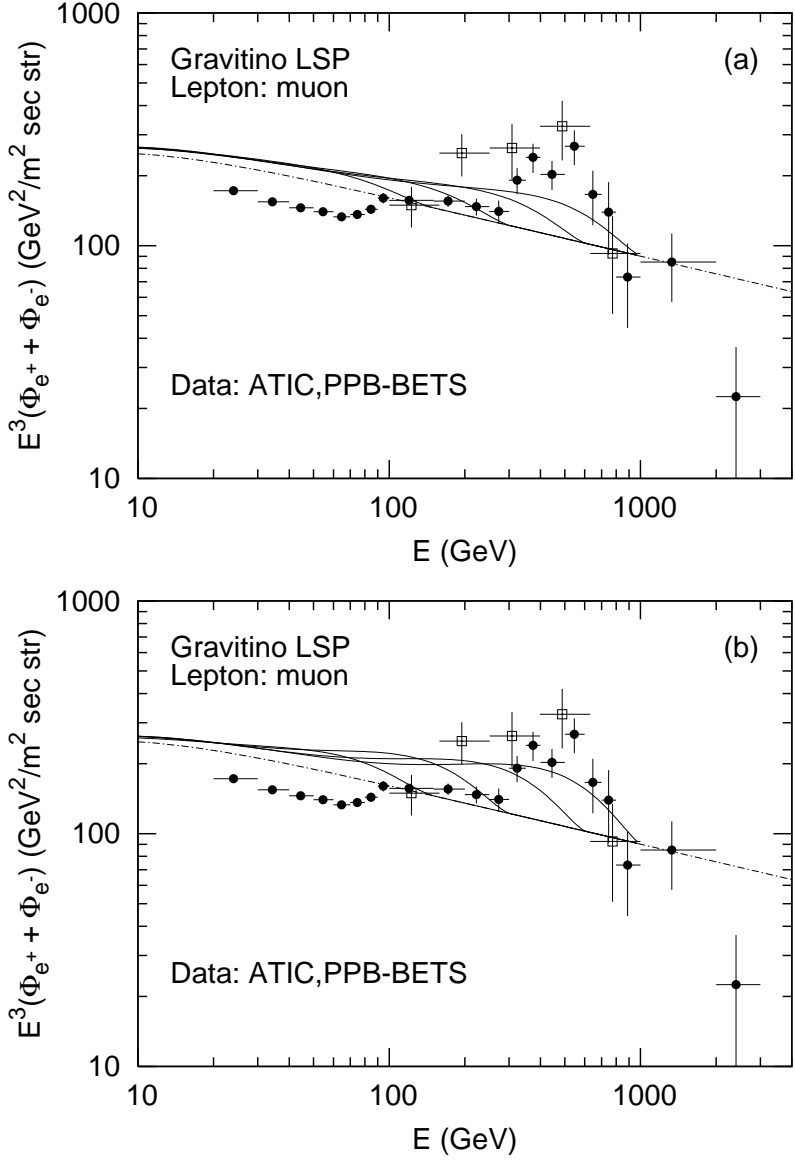}
      \caption{\small Same as Fig.\ \ref{fig:positronflux_e}, but the
        gravitino dominantly decays to second-generation lepton.  We
        take the the same mass and lifetime as
        Fig.\ \ref{fig:positronfraction_mu}, and also take $m_{3/2}=2$
        TeV with $\tau_{3/2}=7.4 \times 10^{25}$ sec and $4.6 \times
        10^{25}$ sec in (a) and (b), respectively.  }
      \label{fig:positronflux_mu}
    \end{center}
    \vspace{-0.5cm}
\end{figure}

\begin{figure}
    \begin{center}
      \includegraphics[scale=1.1]{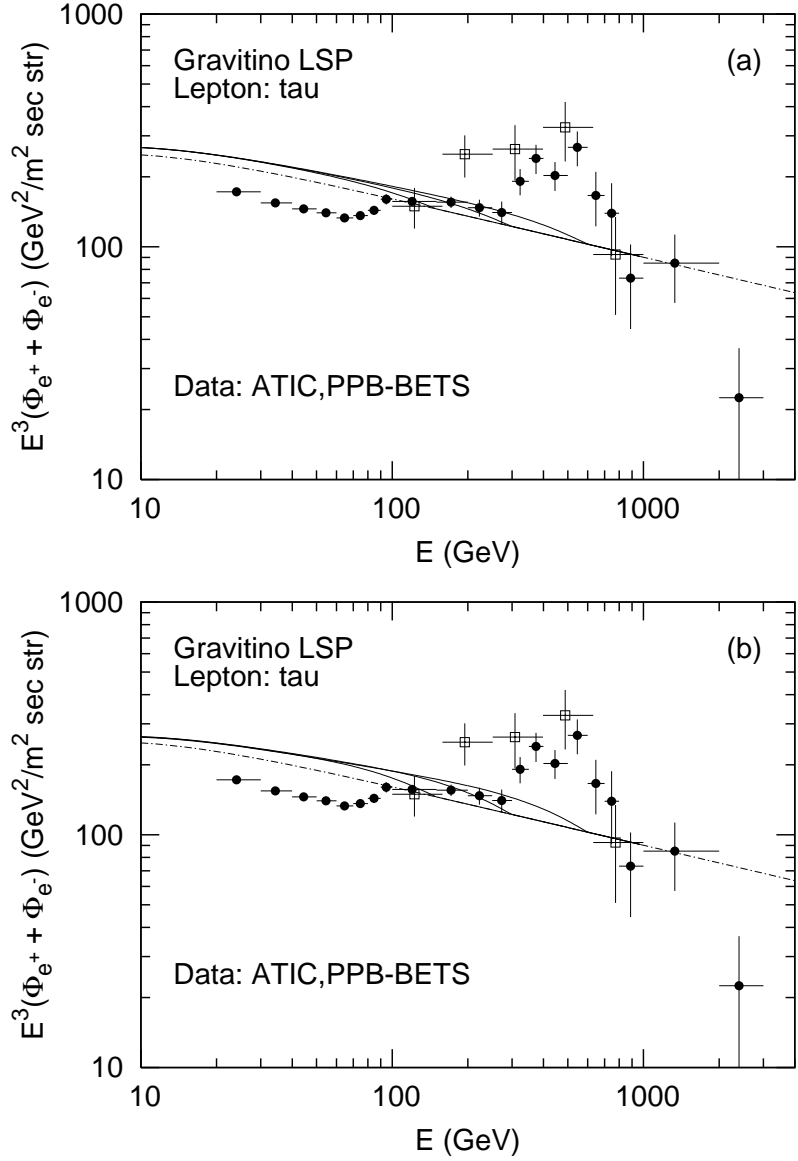}
      \caption{\small Same as Fig.\ \ref{fig:positronflux_e}, but the
        gravitino dominantly decays to third-generation lepton.  We
        take the the same mass and lifetime as
        Fig.\ \ref{fig:positronfraction_tau}.  }
      \label{fig:positronflux_tau}
    \end{center}
    \vspace{-0.5cm}
\end{figure}

Next, we consider the total flux $\Phi_{e^+}+\Phi_{e^-}$.  The
numerical results are shown in Figs.\ \ref{fig:positronflux_e} $-$
\ref{fig:positronflux_tau} for the cases where the gravitino decays
only into first-, second-, or third-generation lepton.  For the
calculation of $\Phi_{e^+}+\Phi_{e^-}$, we use the best-fit lifetime
to fit the PAMELA data.  In the figures, it can be seen that the
observed anomalous structure is well reproduced with both MED and M2
models by relevantly choosing $m_{3/2}$, except for the case that
final-state lepton is third-generation.  For the case of final-state
lepton being in the first- (second-) generation, the result is a good
agreement with the observation when $m_{3/2} \simeq 1.2$ TeV (2 TeV).
Here, we note that the total flux is not sensitive to the background
because the signal from the dark matter is larger than (or at least
comparable to) the background.

With the decay of the gravitino dark matter, energetic $\gamma$-ray
and anti-proton are also produced.  Thus, with the observations of the
fluxes of these particles, we may confirm or exclude the present
scenario.\footnote
{High energy $\gamma$-ray from the Galactic center has been
calculated in the scenario where dark matter annihilates into $W^+W^-$
pair \cite{Bertone:2008xr,Nardi:2008ix}.  We have checked that the
high energy $\gamma$-ray flux in the present scenario is much smaller
than that in the annihilating scenario (into $W^+W^-$ pair), assuming
that the PAMELA and ATIC anomalies are from the decay or the
annihilation of dark matter.  Then, we found that the total
$\gamma$-ray flux in the present scenario is consistent with the HESS
observation \cite{Aharonian:2006wh}.  }
Notice that $\gamma$ and $\bar{p}$ are mostly from the hadronic decays
of the weak and Higgs bosons.  Thus, the fluxes of these particles are
insensitive to the flavor of the final-state lepton, except for
the case where $\tau$-lepton is produced by the decay. (See the
following discussion.)

\begin{figure}[t]
    \begin{center}
      \includegraphics[scale=1.4]{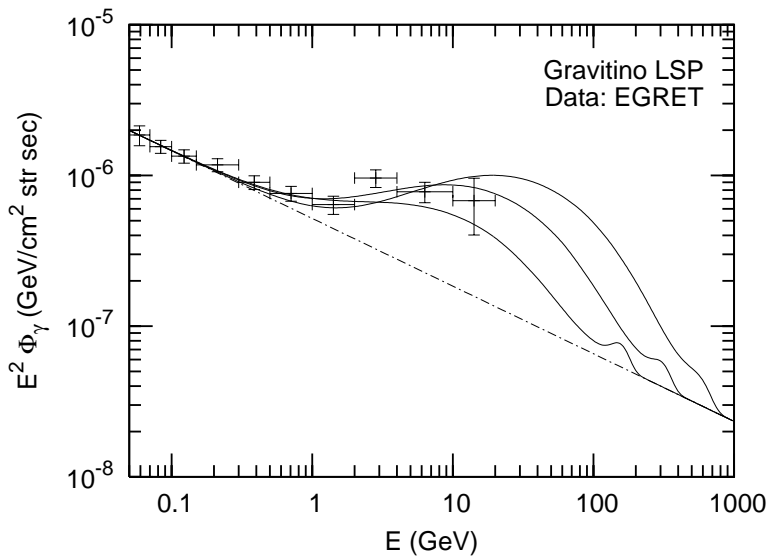}
      \caption{\small $\gamma$-ray flux.  Dot-dashed line is the
     background flux.  Here, we take mass and lifetime as the same as
     figure (a) in Fig.\ \ref{fig:positronfraction_e} }
     \label{fig:gammaray}
    \end{center}
    \vspace{-0.5cm}
\end{figure}

\begin{figure}[t]
    \begin{center}
      \includegraphics[scale=1.4]{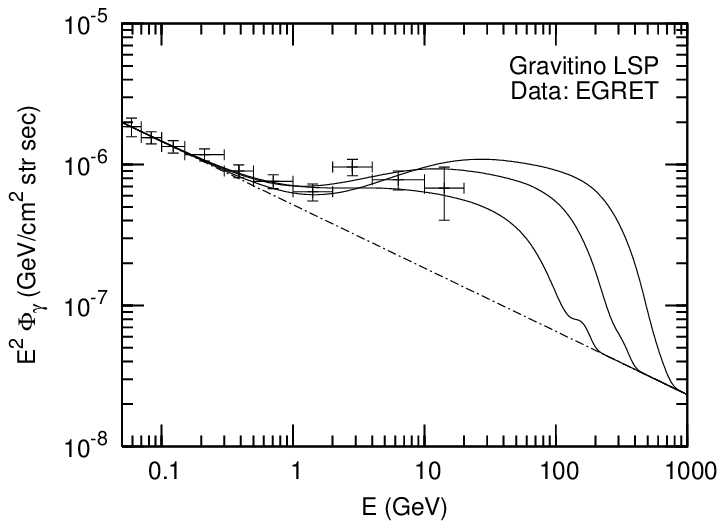}
      \caption{\small The same as Fig. \ref{fig:gammaray}, but for the
     case where primary lepton produced in the decay is
     third-genaration.  } 
     \label{fig:gammaray_tau}
    \end{center}
    \vspace{-0.5cm}
\end{figure}

In Fig.\ \ref{fig:gammaray}, we plot the flux of the $\gamma$-ray in
the present scenario with the primary lepton being in first- or
  second-generation.  From the figure, we see that the expected
$\gamma$-ray flux well agrees with EGRET data irrespective of
$m_{3/2}$.  Because the flux from the gravitino decay is larger than
the background, the $\gamma$-ray flux does not highly depend on the
background.\footnote
{The recent preliminary results of the FERMI satellite (for the
    region $10^{\circ}<b<20^{\circ}$) indicates no anomalous excess in
    the high energy $\gamma$-ray \cite{TITECH_Workshop}.  We have also
    calculated the $\gamma$-ray flux for the region
    $10^{\circ}<b<20^{\circ}$ in the present scenario, and found that
    the dark-matter contribution to the flux for such a region is
    smaller than the FERMI data.}
On the other hand, if the primary lepton produced in the decay is in
  third-genaration, $\gamma$ from $\pi^0$ decay also contributes to the
  total flux \cite{Colafrancesco:2005ji}.  In Fig.\
  \ref{fig:gammaray_tau}, we show the result for such a case.  From the
  figure, one can see an increase of the flux in the high energy region
  of $E \gtrsim $ 100 GeV.

\begin{figure}[t]
    \begin{center}
      \includegraphics[scale=1.4]{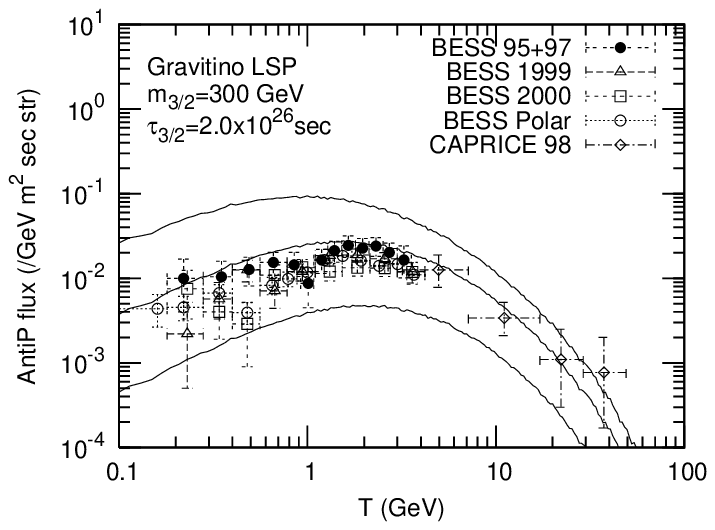}
      \caption{\small Anti-proton flux in MIN, MED, and MAX models.
        Here, we take $m_{3/2}=300~{\rm GeV}$ and $\tau_{3/2}=2.0
        \times 10^{26}~{\rm sec}$, which is the best-fit lifetime in
        the case that the gravitino dominantly decays to
        first-generation lepton, and also plot the observation data by
        BESS \cite{Orito:1999re} and CAPRICE \cite{Boezio:2001ac}. }
      \label{fig:anti-pfluxMg300}
    \end{center}
    \vspace{-0.5cm}
\end{figure}

\begin{figure}
    \begin{center}
      \includegraphics[scale=1.4]{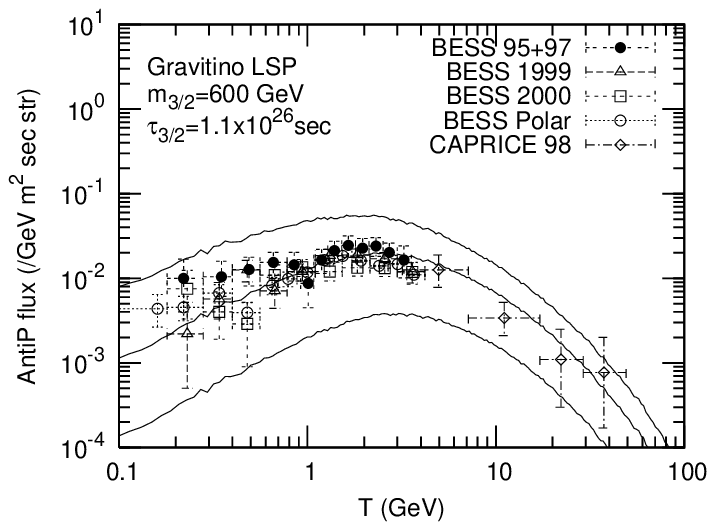}
      \caption{\small Same as Fig.\ \ref{fig:anti-pfluxMg300},
        except for taking $m_{3/2}=600$ GeV and corresponding best-fit
        lifetime.  }
      \label{fig:anti-pfluxMg600}
    \end{center}
    \vspace{-0.5cm}
\end{figure}

\begin{figure}
    \begin{center}
      \includegraphics[scale=1.4]{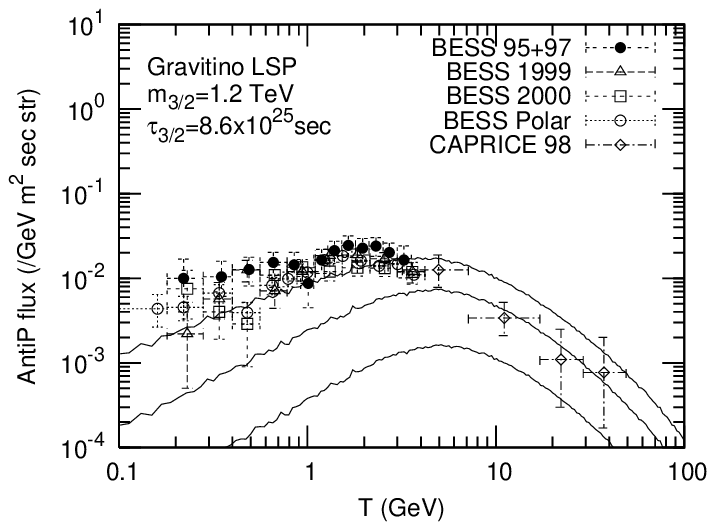}
      \caption{\small Same as Fig.\ \ref{fig:anti-pfluxMg300},
        except for taking $m_{3/2}=1.2$ TeV and corresponding best-fit
        lifetime.  }
      \label{fig:anti-pfluxMg1200}
    \end{center}
    \vspace{-0.5cm}
\end{figure}

The anti-proton flux from the decay of the gravitino dark matter is
shown in Figs.\ \ref{fig:anti-pfluxMg300} $-$
\ref{fig:anti-pfluxMg1200}.  We can see that the anti-proton flux
depends on the propagation model.  However, with the MED and MIN
models, the flux from the gravitino decay is comparable to or smaller
than the observed fluxes.\footnote
{We also estimated anti-proton to proton ratio to compare recent data by
    PAMELA \cite{Adriani:2008zq}, using the proton flux observed by BESS
    \cite{Sanuki:2000wh} and CAPRICE \cite{Boezio:1998vc}.  Then we
    found that the $\bar{p}/p$ ratio is order of magnitude smaller than
    the PAMELA data if we take the MIN propagation model and that it is
    comparable to or a few times larger than the PAMELA data with MED
    model.  Thus, taking account account of the uncertainties in the
    propagation model as well as those in the background proton flux,
    the present scenario is not excluded yet.}
Thus, we conclude that the present scenario
is not excluded by the observation of the cosmic-ray $\bar{p}$ flux,
taking account of the uncertainties in the propagation model and
estimation of the background.  However, if a better understanding of
the propagation of the anti-proton becomes available in the future, it
will provide a significant test of the present scenario.

Before closing this section, we comment that this scenario may be
tested by the LHC experiment.  Indeed, in this scenario, the lightest
superparticle in the MSSM sector (which we call MSSM-LSP) may decay
inside the detector.  In the present scenario, the MSSM-LSP is likely
to decay to the standard model particles via the RPV interaction even
though there exists a superparticle (i.e., gravitino) lighter than the
MSSM-LSP.  As we have mentioned, the $\kappa$ parameter is expected to
be $O(10^{-10})$, which gives the lifetime of the MSSM-LSP of the
order of $\sim 10^{-(4-5)}\ {\rm sec}$.  Thus, the typical decay
length of the MSSM-LSP is expected to be much longer than the sizes of
the ATLAS and CMS detectors.  However, since enormous amount of SUSY
events is expected at the LHC experiment, some of the MSSM-LSP
produced at the LHC may decay inside the detector, which results in
drastic signal, like a displaced vertex or a kink in a high-energy
charged track.  In addition, if such decay processes can be observed,
it may be also possible to constrain the lifetime of the MSSM-LSP,
which may be used for the determination of the $\kappa$ parameter
\cite{Ishiwata:2008tp}.

\section{Neutralino LSP}
\label{sec:neutralino}
\setcounter{equation}{0}

In the previous section, we have considered the case that the
gravitino is the LSP.  In conventional SUSY models, another important
candidate for the LSP is the lightest neutralino, which is a linear
combination of Bino, neutral Wino, and two neutral Higgsinos.  To make
our discussion simple, in this Section, we assume that the lightest
neutralino is (almost) Bino-like.  This is the case, for example, in
large fraction of the parameter space of models with the
grand-unification condition among gaugino masses.

With the RPV operators given in Eq.\ \eqref{L_RPV}, the Bino-like
neutralino $\tilde{B}$ decays as $\tilde{B}\rightarrow Z\nu_i$,
$Wl_i$, and $h\nu_i$.  Decay rates for these processes are given
by\footnote
{Here, we assume that the lightest Higgs boson is almost
  standard-model like.}
\begin{eqnarray}
  \Gamma_{\tilde{B} \rightarrow Z \nu_i}
  &=&
  \frac{1}{128 \pi}g^2_Z \sin^2 \theta_W \kappa^2_i \ 
  m_{\tilde{B}}
  \left(
    1-3\frac{m^4_Z}{m^4_{\tilde{B}}}+2\frac{m^6_Z}{m^6_{\tilde{B}}}
  \right),
  \label{Bino2Znu}
  \\
  \Gamma_{\tilde{B} \rightarrow W l_i}
  &=&
  \frac{1}{64 \pi} g^2_Z \sin^2 \theta_W \kappa^2_i \ 
  m_{\tilde{B}}
  \left(
    1-3\frac{m^4_W}{m^4_{\tilde{B}}}+2\frac{m^6_W}{m^6_{\tilde{B}}}
  \right),
  \label{Bino2Wl}
  \\
  \Gamma_{\tilde{B} \rightarrow h \nu_i}
  &=&
  \frac{1}{128 \pi} g^2_Z \sin^2 \theta_W \kappa^2_i \ 
  m_{\tilde{B}}
  \left(\frac{m^2_{\tilde{\nu}}}{m^2_{\tilde{\nu}}-m^2_h}
  \right)^2
  \left(
    1-\frac{m^2_h}{m^2_{\tilde{B}}}
  \right)^2,
  \label{Bino2Hnu}
\end{eqnarray}
where $g_Z=\sqrt{g_1^2+g_2^2}$ (with $g_1$ and $g_2$ being the gauge
coupling constants of the $U(1)_Y$ and $SU(2)_L$ gauge groups,
respectively), $\theta_W$ is the Weinberg angle, $m_{\tilde{B}}$ is
Bino-like neutralino mass, and $m_Z$, $m_W$, $m_h$ are masses of
corresponding gauge and Higgs bosons.

As one can see from the above decay rates, we obtain the relation
$\Gamma_{\tilde{B} \rightarrow Z \nu_i}\simeq \Gamma_{\tilde{B}
  \rightarrow h \nu_i}\simeq\frac{1}{2}\Gamma_{\tilde{B} \rightarrow W
  l_i}$.  Remember that, for the case of the gravitino LSP, the same
(approximated) relation holds.  Thus, the fluxes of the high energy
cosmic rays in the Bino LSP case is expected to be similar to that in
the gravitino LSP case as far as the lifetime is $\sim 10^{26}\ {\rm
  sec}$.  Since the decay rate of the Bino is not suppressed by the
Planck scale, the size of the $\kappa_i$ parameter relevant to explain
the PAMELA and ATIC anomalies is much smaller than that in the
gravitino LSP case.  Indeed, with the decay rates given in Eq.\
\eqref{Bino2Znu} $-$ \eqref{Bino2Hnu}, the lifetime of the Bino LSP in
the present case is estimated as
\begin{eqnarray}
  \tau_{\tilde{B}} \simeq
  2 \times 10^{25}\ {\rm sec} \times
  \left( \frac{\kappa}{10^{-25}} \right)^{-2} 
  \left( \frac{m_{\tilde{B}}}{1\ {\rm TeV}} \right)^{-1}.
\end{eqnarray}


So far, we have considered the bi-linear RPV interaction given in Eq.\
\eqref{L_RPV}.  In such a case, as we have discussed, high energy
$\gamma$ and anti-proton are also produced by the decay of the LSP.
However, if we consider other types of RPV interactions, it may be
possible to enhance the $e^+$ and $e^-$ fluxes without affecting
$\gamma$-ray and anti-proton fluxes.  This is the case where the Bino
LSP decays mainly via the PRV superpotential:
\begin{eqnarray}
  W_{\rm RPV} = \frac{1}{2} \lambda_{ijk} 
  \hat{L}_i \hat{L}_j \hat{E}^c_k,
  \label{W_LLE}
\end{eqnarray}
where $\lambda_{ijk}=-\lambda_{jik}$.  With this superpotential, the
Bino decays as $\tilde{B}\rightarrow \nu_i l_{L,j}^\pm l_{R,k}^\mp$
and $\nu_j l_{L,i}^\pm l_{R,k}^\mp$ via diagrams with a (virtual)
slepton propagation.  Hereafter, let us consider the $e^+$ and $e^-$
fluxes in such a case.

With the above superpotential, the Bino decays into the three-body
final state, and hence the final-state leptons are not monochromatic.
For simplicity, we consider the case that the right-handed sleptons
are lighter than left-handed ones, so that the diagram with the
propagator of the right-handed slepton dominantly contributes to the
Bino decay.  Then, denoting the energies of $l_{L,j}^\pm$ and
$l_{R,k}^\mp$ in the rest frame of $\tilde{B}$ as $E_{l_L}$ and
$E_{l_R}$, respectively, the energy distribution of the charged
leptons for the process $\tilde{B}\rightarrow \nu_i l_{L,j}^\pm
l_{R,k}^\mp$ is given by
\begin{eqnarray}
  \frac{d\Gamma_{\tilde{B}\rightarrow \nu_i l_{L,j}^\pm l_{R,k}^\mp}}
  {d E_{l_L} d E_{l_R}}
  = \frac{g_1^2 \lambda_{ijk}^2}{64 \pi^3 m_{\tilde{B}}}
  \frac{z_{l_R} (1-z_{l_R})}
       {[(m_{\tilde{l}_{R,k}}/m_{\tilde{B}})^2 - 1 + z_{l_R}]^2},
\end{eqnarray}
Where $z_{l_{L,R}}\equiv 2E_{l_{L,R}}/m_{\tilde{B}}$.  (Notice that
$0\leq z_{l_{L,R}}\leq 1$.)

\begin{figure}[t]
    \begin{center}
      \includegraphics[scale=1.1]{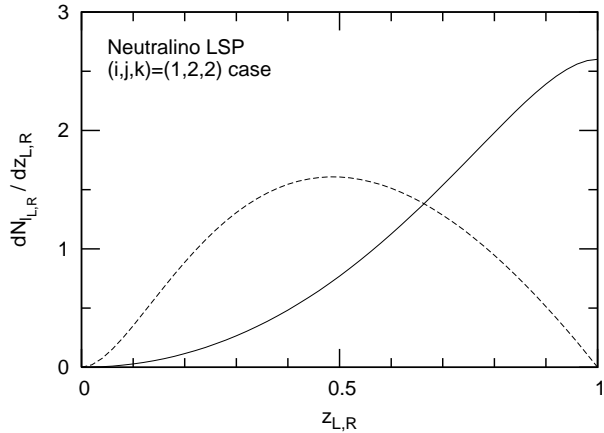}
      \caption{\small Distribution of the left- (solid) and
        right-handed (dashed) charged leptons emitted from the Bino
        decay via the $\hat{L}\hat{L}\hat{E}^c$-type superpotential.
	Here, we take $m_{\tilde{l}_{R}}/m_{\tilde{B}}=1.2$.}
      \label{fig:energydist_LR}
    \end{center}
\end{figure}

In Fig.\ \ref{fig:energydist_LR}, we plot the energy distributions of
the final-state leptons:
\begin{eqnarray}
  \frac{d N_{l_{L,R}}}{d z_{L,R}} \equiv
  \frac{1}{\Gamma_{\tilde{B}\rightarrow \nu_i l_{L,j}^\pm l_{R,k}^\mp}}
  \int d z_{l_{R, L}}
  \frac{d\Gamma_{\tilde{B}\rightarrow \nu_i l_{L,j}^\pm l_{R,k}^\mp}}
  {d z_{l_L} d z_{l_R}}.
\end{eqnarray}
Notice that this quantity depends only on the ratio
$m_{\tilde{l}_{R,k}}/m_{\tilde{B}}$. (In the figure, we take
$m_{\tilde{l}_{R}}/m_{\tilde{B}}=1.2$.)  As one can see, the
left-handed lepton emitted in the decay is likely to be energetic.
Thus, if the coupling constant $\lambda_{1jk}$ is sizable so that the
Bino decays dominantly as $\tilde{B}\rightarrow e_{L}^\pm \nu_{j}
l_{R,k}^\mp$, significant effects on the electron and positron fluxes
are expected.

\begin{figure}
    \begin{center}
      \includegraphics[scale=1.1]{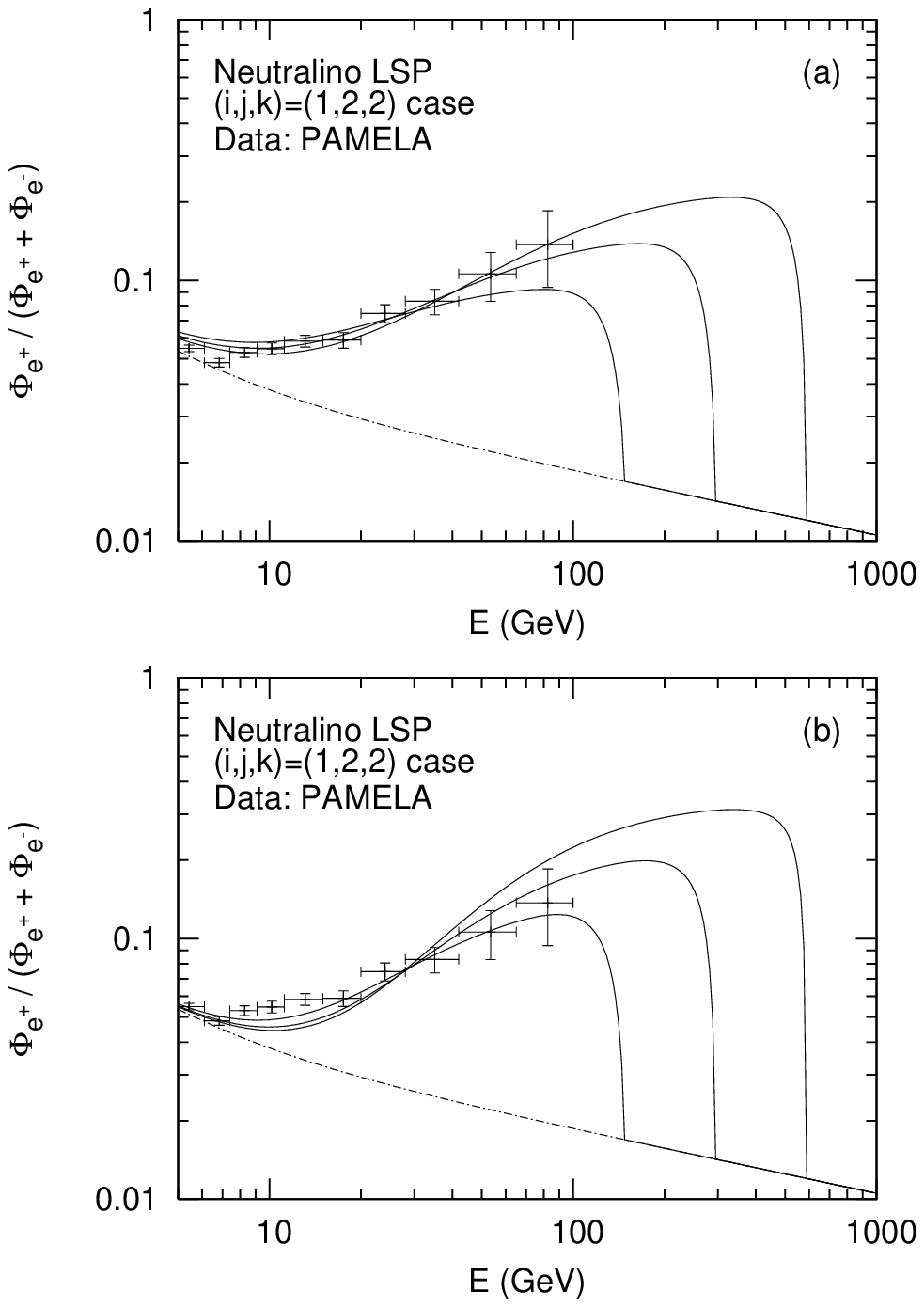}
      \caption{\small Positron fractions in (a) MED and (b) M2 models
        for the case where only $\lambda_{122}$ is non-zero.  Here, we
        take $m_{\tilde{B}}=300\ {\rm GeV}$, 600 GeV, and 1.2 TeV from
        left to right and $\tau_{\tilde{B}}=3.2 \times 10^{26}\ {\rm
          sec}$, $2.0 \times 10^{26}~{\rm sec}$, and $1.2 \times
        10^{26}$ sec ($2.0\ \times 10^{26}\ {\rm sec}$, $1.1\ \times
        10^{26}\ {\rm sec}$, and $5.0\ \times 10^{25}\ {\rm sec}$) in
        MED (M2) model, respectively.  }
      \label{fig:positronfraction_bino122}
    \end{center}
\end{figure}

\begin{figure}
    \begin{center}
      \includegraphics[scale=1.1]{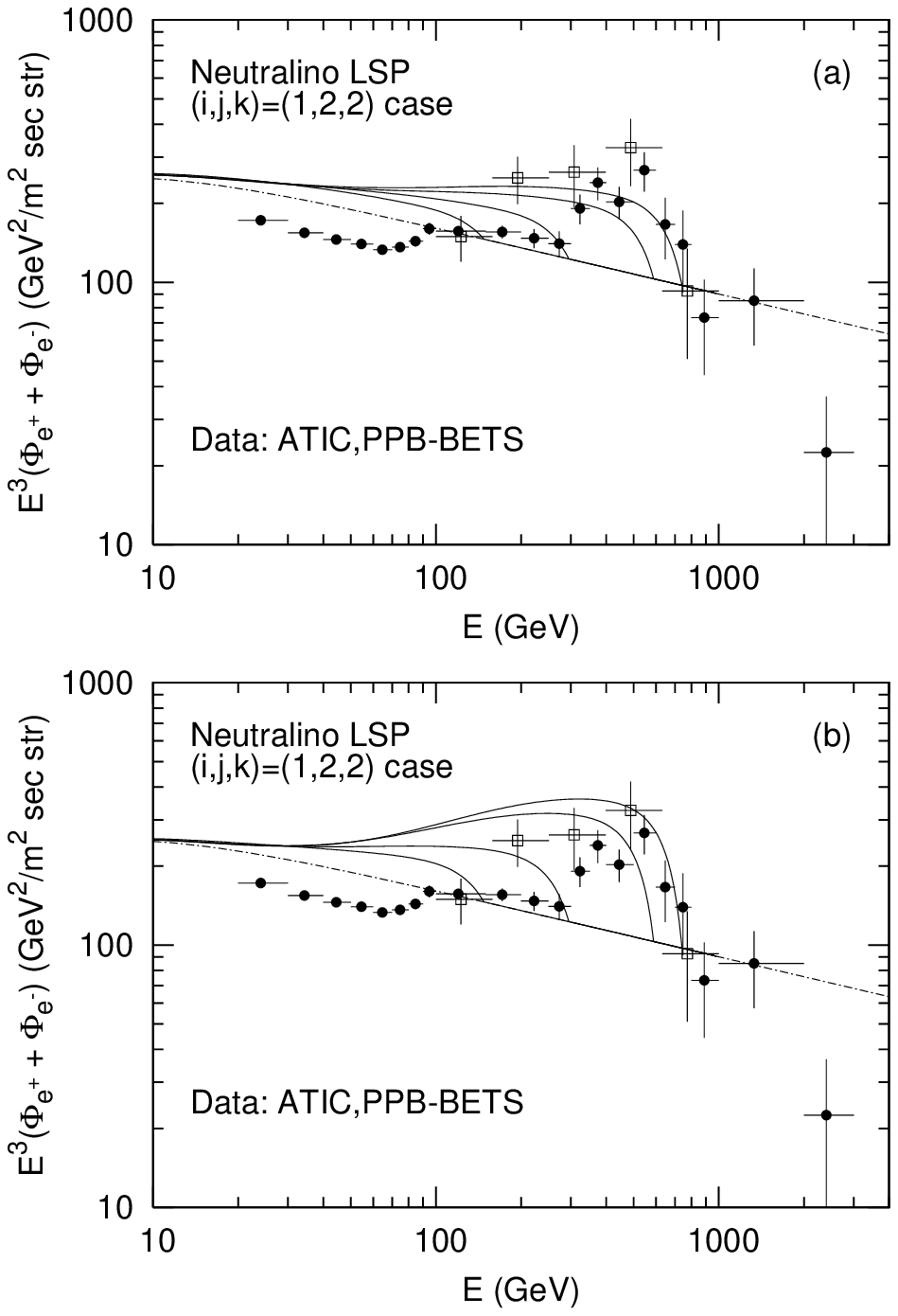}
      \caption{\small Total fluxes in (a) MED and (b) M2 models for
        the case where only $\lambda_{122}$ is non-zero.  We take the
        same values for $m_{\tilde{B}}$ and $\tau_{\tilde{B}}$ in
        Fig.\ \ref{fig:positronfraction_bino122}, and also take
        $m_{\tilde{B}}=1.5$ TeV with $\tau_{\tilde{B}}=9.3 \times
        10^{25}$ sec ($4.0 \times 10^{25}$ sec) in (a)((b)).}
      \label{fig:positronflux_bino122}
    \end{center}
\end{figure}

In Fig.\ \ref{fig:positronfraction_bino122}, we show the positron
fraction in the Bino dark matter case with the RPV superpotential
given in Eq.\ \eqref{W_LLE}.  Here, we consider the case that
$\lambda_{122}$ is the largest so that the Bino decays via this
coupling, and we use $m_{\tilde{l}_{R}}/m_{\tilde{B}}=1.2$.  The
best-fit lifetime to explain the PAMELA anomaly depends on
$m_{\tilde{B}}$ as well as on the propagation model as in the case of
the gravitino LSP.  For the MED (M2) model of propagation, the
best-fit lifetime is given by $\tau_{3/2}=3.2 \times 10^{26}\ {\rm
sec}$, $2.0 \times 10^{26}~{\rm sec}$, and $1.2 \times 10^{25}$ sec
($2.0\ \times 10^{26}\ {\rm sec}$, $1.1\ \times 10^{26}\ {\rm sec}$,
and $5.0\ \times 10^{25}\ {\rm sec}$) for $m_{\tilde{B}}=300\ {\rm
GeV}$, $600\ {\rm GeV}$, and $1.2\ {\rm TeV}$, respectively.  In
addition, in Fig.\ \ref{fig:positronflux_bino122}, we plot the flux of
$e^++e^-$.  We can see that the simultaneous explanation of the
PAMELA and ATIC/PPB-BETS anomalies may be possible in the present
scenario if $m_{\tilde{B}}\simeq 1.5\ {\rm TeV}$ and
$\tau_{\tilde{B}}\simeq 9.3 \times 10^{25} {\rm sec}$ ($4.0 \times
10^{25}$ sec) for MED (M2) model.

So far, we have considered the case that $\lambda_{122}$ is the
largest.  We have checked that similar results for the positron
fraction and the total flux are obtained as far as the
first-generation lepton is emitted in the decay.  If $\tilde{B}$
decays only into second- and third-generation leptons, on the
contrary, the total flux $\Phi_{e^+}+\Phi_{e^-}$ is suppressed
compared to the observed values when we adopt the best-fit lifetime to
explain the PAMELA anomaly.

\section{Sneutrino LSP}
\label{sec:sneutrino}
\setcounter{equation}{0}

The third candidate for the LSP is the sneutrino.  In the framework of
the MSSM, the sneutrino may not be a popular candidate for the LSP.
However, the lightest sneutrino can be dark matter without conflicting
phenomenological constraints, as we discuss below.  In the previous
section, we have seen that, with the $\hat{L}\hat{L}\hat{E}^c$ type
RPV superpotential given in Eq.\ \eqref{W_LLE}, enhancement of
electron and positron fluxes is possible without affecting the
$\gamma$-ray and anti-proton fluxes.  If the Bino is the LSP, the
final-state leptons are not monochromatic, so that the end-point
behavior of the cosmic-ray $e^\pm$ is smoothed.  On the other hand, if
the sneutrino is the LSP and hence is dark matter, and also if it
decays via the $\hat{L}\hat{L}\hat{E}^c$ type RPV superpotential,
shape of the $e^\pm$ spectrum drastically changes and a sharp edge at
the end-point can be obtained \cite{Ishiwata:2008cv}.

In the MSSM, there only exist three left-handed sneutrinos
$\tilde{\nu}_L$s (``$L$'' means left-handed in this Section), and one of
them may be the lightest superparticle.  By tuning the soft
SUSY-breaking scalar and gaugino masses, the sneutrino becomes the LSP
in some parameter region.  In addition, the direct detection
constraint on the sneutrino dark matter \cite{Falk:1994es} can be
avoided by introducing small lepton-number violating operator $\sim
(\tilde{L}H_u)^2$ \cite{Hall:1997ah}.  In addition, there is another
possibility to realize the sneutrino dark matter scenario, in which
the right-handed sneutrino $\tilde{\nu}_R$ becomes the LSP
\cite{SnurDM}.  Since various neutrino-oscillation experiments suggest
that the neutrinos are massive, we expect the existence of the
right-handed (s)neutrinos.  If the neutrino masses are Dirac-type,
$\tilde{\nu}_R$ becomes as light as other MSSM superparticles in the
gravity-mediation SUSY breaking scenario.  (In other cases,
$\tilde{\nu}_R$ may be much lighter than MSSM superparticles.)  Thus,
in some parameter space, $\tilde{\nu}_R$ can be the LSP.  In such a
case, $\tilde{\nu}_R$ can also be dark matter.  The following
arguments holds if $\tilde{\nu}_L$ or $\tilde{\nu}_R$ is the LSP.

If the sneutrino in $i$-th generation is the LSP, it decays as
$\tilde{\nu}_i\rightarrow l^+_j l^-_k$ assuming that the dominant RPV
interaction is given by the $\hat{L}\hat{L}\hat{E}^c$ type RPV
superpotential given in Eq.\ \eqref{W_LLE}.  The decay rate of this
process is
\begin{eqnarray}
  \Gamma_{\tilde{\nu}_i\rightarrow l^+_j l^-_k} = 
  \frac{\lambda_{ijk}^2 \theta_{\tilde{\nu}}^2}{16 \pi}
  m_{\tilde{\nu}_i},
  \label{Gamma(snu)}
\end{eqnarray}
where $\theta_{\tilde{\nu}}$ is the mixing parameter in the sneutrino
sector.  When the $\tilde{\nu}_L$ is the LSP,
$\theta_{\tilde{\nu}}=1$.  Even if $\tilde{\nu}_i$ is right-handed, it
decays via the RPV interaction given in Eq.\ \eqref{W_LLE} because
there should exist left-right mixing term of the sneutrinos, which is
of the form
\begin{eqnarray}
  {\cal L}_{LR} = \delta m_{\tilde{\nu}_L \tilde{\nu}_R}^2
  \tilde{\nu}_L \tilde{\nu}_R + {\rm h.c.}
\end{eqnarray}
In such a case, $\theta_{\tilde{\nu}}$ in Eq.\ \eqref{Gamma(snu)}
should be
\begin{eqnarray}
  \theta_{\tilde{\nu}} = \frac{\delta m_{\tilde{\nu}_L \tilde{\nu}_R}^2}
  {m_{\tilde{\nu}_L}^2 - m_{\tilde{\nu}_R}^2}.
\end{eqnarray}
Thus, in any case, the sneutrino LSP decays into charged-lepton pair
via the RPV interaction given in Eq.\ \eqref{W_LLE}.  Thus, if the
sneutrino is dark matter, those leptons become the source of
cosmic-ray $e^\pm$.  Importantly, in such a case, dark matter decays
only leptonically, and the productions of hadrons and photon are
irrelevant.  Thus, the present model may produce significant amount of
$e^\pm$ in cosmic ray without affecting the anti-proton and
$\gamma$-ray fluxes.

In the following, we calculate the electron and positron fluxes from
the decay of sneutrino dark matter.  Here, we take $m_{\tilde{\nu}}=
300\ {\rm GeV}$, $600\ {\rm GeV}$, and $1.2\ {\rm TeV}$.  The
resultant flux depends on the flavors of the final-state leptons.
Here, for simplicity, we consider three simple decay modes:
$\tilde{\nu}\rightarrow e^+e^-$, $\mu^+\mu^-$, and $\tau^+\tau^-$.

\begin{figure}
    \begin{center}
      \includegraphics[scale=1.4]{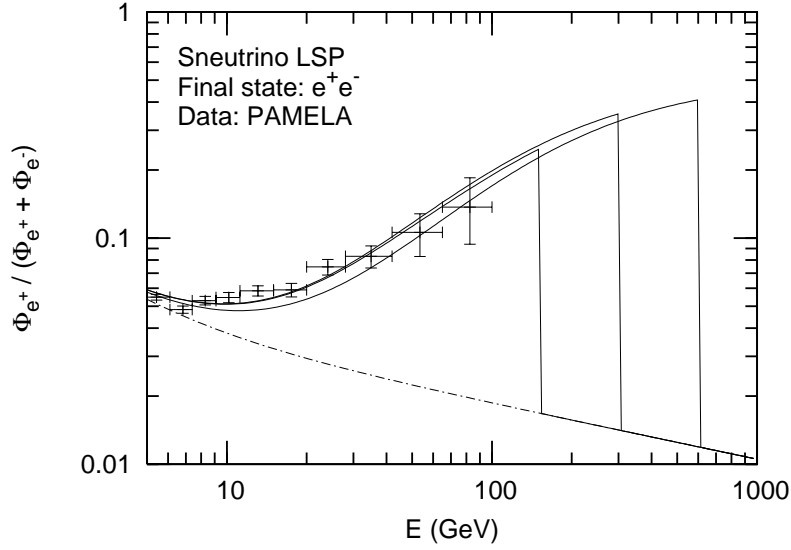}
      \caption{\small Positron fractions in MED model for the case
        where the sneutrino decays to the final state: $e^+e^-$.
        Here, we take $m_{\tilde{\nu}}=300\ {\rm GeV}$, 600 GeV, and
        1.2 TeV from left to right and $\tau_{\tilde{\nu}}=5.4 \times
        10^{26}\ {\rm sec}$, $2.5 \times 10^{26}~{\rm sec}$, and $1.6
        \times 10^{26}$ sec.  }
      \label{fig:positronfraction_ee}
    \end{center}
\end{figure}

\begin{figure}
    \begin{center}
      \includegraphics[scale=1.4]{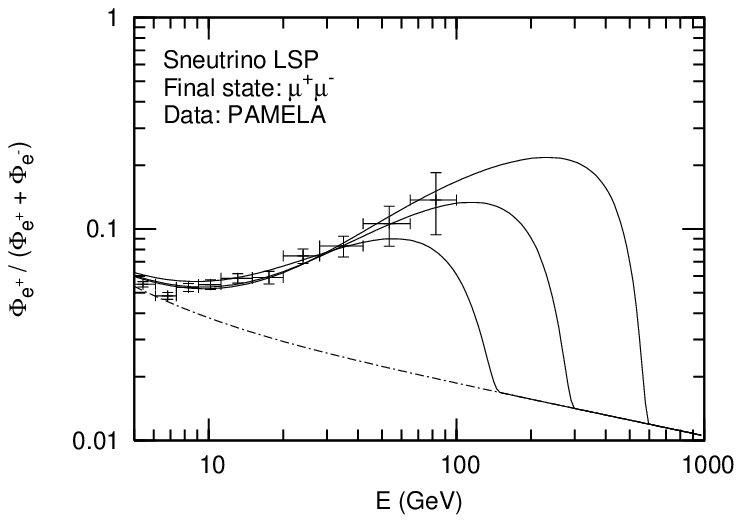}
      \caption{\small Same as Fig.\ \ref{fig:positronfraction_ee} in
        the case of final state: $\mu^+\mu^-$.  Here, we take
        $\tau_{\tilde{\nu}}=3.7 \times 10^{26}\ {\rm sec}$, $2.3
        \times 10^{26}~{\rm sec}$, and $1.2 \times 10^{26}$ sec, 
        which are the best-fit lifetime with MED model.  }
      \label{fig:positronfraction_mumu}
    \end{center}
\end{figure}

\begin{figure}
    \begin{center}
      \includegraphics[scale=1.4]{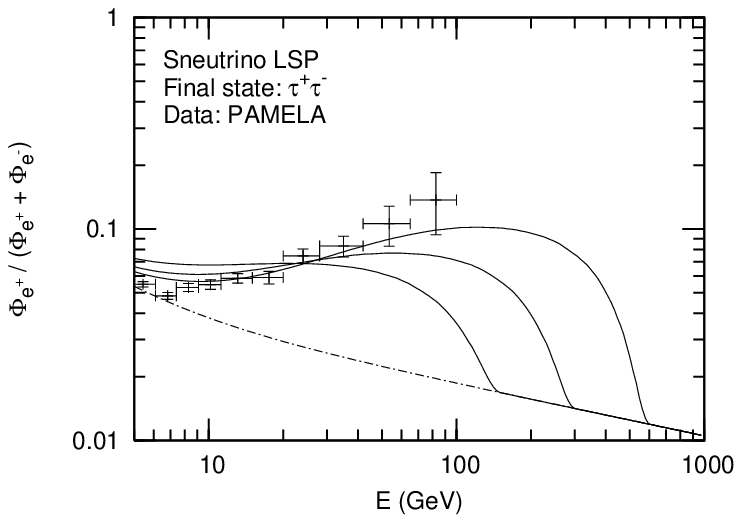}
      \caption{\small Same as Fig.\ \ref{fig:positronfraction_ee} in
        the case of final state: $\tau^+\tau^-$.  Here, we take
        $\tau_{\tilde{\nu}}=1.8 \times 10^{26}\ {\rm sec}$, $1.5
        \times 10^{26}~{\rm sec}$, and $1.1 \times 10^{26}$ sec,
        which are the best-fit lifetime with MED model.  }
      \label{fig:positronfraction_tautau}
    \end{center}
\end{figure}

The results of the calculations of positron fraction are shown in
Fig.\ \ref{fig:positronfraction_ee} $-$
\ref{fig:positronfraction_tautau}, where MED mode of the propagation
is adopted.  In addition, we use the best-fit lifetime for each
$m_{\tilde{\nu}}$.  We see good agreements with the observation for all
the cases.  We have also checked that a reasonable agreement between
the prediction and the PAMELA data is obtained even with the M2 model
if the sneutrino decays into $\mu^+\mu^-$ and $\tau^+\tau^-$.

\begin{figure}[t]
    \begin{center}
      \includegraphics[scale=1.4]{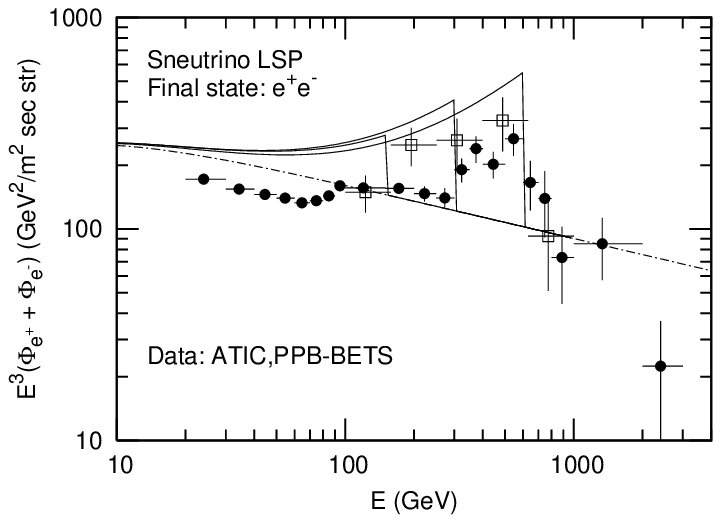}
      \caption{\small Total fluxes of positron and electron in MED
        model for the case sneutrino decays to the final state:
        $e^+e^-$.  Here, we take the the same mass and lifetime as
        Fig.\ \ref{fig:positronfraction_ee}, and also take
        $m_{\tilde{\nu}}=1.5$ TeV with $1.2 \times 10^{26}$ sec.
      }
      \label{fig:positronflux_ee}
    \end{center}
\end{figure}

\begin{figure}
    \begin{center}
      \includegraphics[scale=1.4]{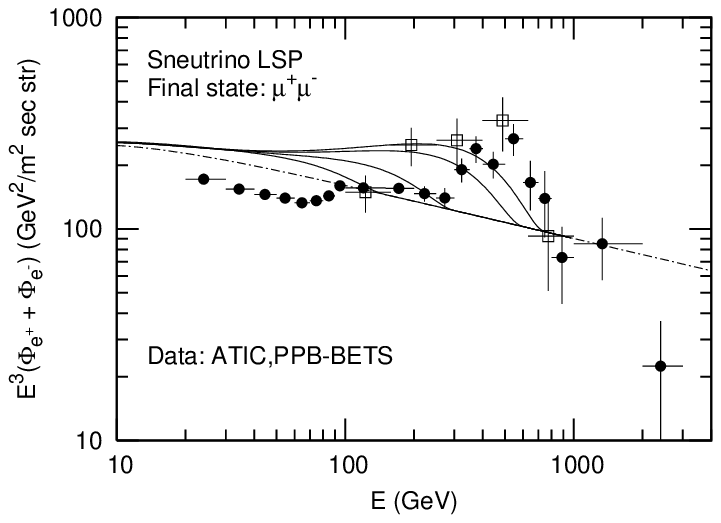}
      \caption{\small Same as Fig.\ \ref{fig:positronflux_ee} for the
        case of final state: $\mu^+\mu^-$.  We take the the same mass
        and lifetime as Fig.\ \ref{fig:positronfraction_mumu}.  }
      \label{fig:positronflux_mumu}
    \end{center}
\end{figure}

\begin{figure}
    \begin{center}
      \includegraphics[scale=1.4]{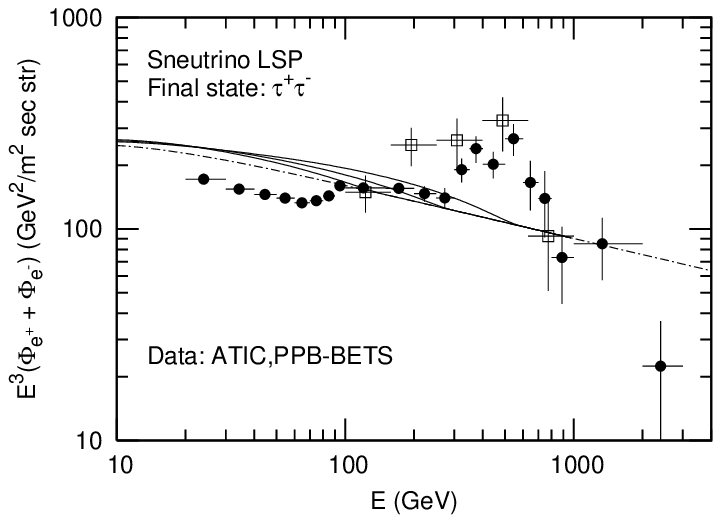}
      \caption{\small Same as Fig.\ \ref{fig:positronflux_ee} for the
        case of final state: $\tau^+\tau^-$.  We take the the same
        mass and lifetime as Fig.\ \ref{fig:positronfraction_tautau}.  }
      \label{fig:positronflux_tautau}
    \end{center}
\end{figure}

Next, we discuss the total flux $\Phi_{e^+}+\Phi_{e^-}$.  The
numerical results are shown in Figs.\ \ref{fig:positronflux_ee} $-$
\ref{fig:positronflux_tautau}, taking the same parameters as in Fig.\
\ref{fig:positronfraction_ee} $-$ \ref{fig:positronfraction_tautau},
respectively.  One can see the anomalous behavior is explained when
$m_{\tilde{\nu}} \sim 1.2-1.5$ TeV in the cases of final state:
$e^+e^-$ and $\mu^+\mu^-$ with MED model.  In addition, we checked
that the same anomalous behavior is obtained in the case of final
state: $\mu^+\mu^-$ with M2 model.

\section{Conclusions and Discussion}
\label{sec:conclusions}
\setcounter{equation}{0}

In this paper, we have studied the cosmic-ray fluxes from the decay of
LSP dark matter, motivated by the recently reported anomalies by
PAMELA and ATIC.  We have introduced several types RPV operators so
that the LSP becomes unstable, and calculated the fluxes of $e^\pm$,
as well as those of $\bar{p}$, and $\gamma$-ray, assuming that the LSP
is the dominant component of dark matter.  The detailed shape of the
spectra of cosmic-ray $e^\pm$ depend on the properties of the LSP dark
matter.  However, when the lifetime of the LSP is of $O(10^{26}\ {\rm
  sec})$, the predicted positron fraction can be consistent with the
observed one by PAMELA irrespective of the mass of the LSP.  On the
contrary, in order to explain the ATIC anomaly for the total flux
$\Phi_{e^+}+\Phi_{e^-}$, the mass of the LSP is required to be $1-1.5\
{\rm TeV}$, depending on the decay modes of the LSP.  If a
monochromatic $e^\pm$ is produced by the two-body decay process, the
LSP mass can be as light as $\sim 1\ {\rm TeV}$, while larger mass is
required for the ATIC anomaly for other cases.  In any case, in order
to explain the ATIC anomaly, relatively large value of the MSSM
particle masses are required, which may make it difficult to solve the
naturalness problem with supersymmetry.

When the LSP decays via the bi-linear RPV interaction given in Eq.\
\eqref{L_RPV}, significant amount of $\gamma$ and anti-proton are also
produced by the decay.  In particular, it has been discussed that the
anti-proton flux may give a stringent constraint on the decaying dark
matter scenario \cite{Ibarra:2008qg}.  However, taking account of the
uncertainties in the propagation models as well as the errors in the
observed fluxes, we have shown that the scenarios discussed in this
paper are not excluded by the present observations.  In addition, we
have also seen that the $\gamma$-ray flux is consistent with the
currently available observational results.  With improved knowledges
about the propagation of the anti-proton, more detailed test of the
scenario may become possible in the future.  It is also notable that a
precise measurement of the cosmic-ray $\gamma$ flux is expected by the
Fermi Telescope.  Furthermore, informations about the decaying dark
matter may be imprinted in the synchrotron radiation from the Galactic
center \cite{Ishiwata:2008qy} and in high-energy neutrino flux
\cite{Hisano:2008ah}.  Future improvements of the observations of
cosmic-ray fluxes should provide better understandings of the
properties of dark matter.

\noindent
{\it Acknowledgments:}
This work was supported in part by Research Fellowships of the Japan
Society for the Promotion of Science for Young Scientists (K.I.), and
by the Grant-in-Aid for Scientific Research from the Ministry of
Education, Science, Sports, and Culture of Japan, No.\ 19540255
(T.M.).
 
\appendix

\section{Green's Function}
\label{appendix:greenFn}
\setcounter{equation}{0} 

We discuss here how we obtain the Green's function for cosmic ray
positrons from the LSP decay. As mentioned in Section \ref{sec:setup},
the Green's function is obtained by solving the diffusion equation
\eqref{DiffEq_epm} with the boundary condition $f_{e^\pm}(E,\vec{x}) =
0$ at the surface of the diffusion zone. Because of the condition, it
is convenient to expand the solution by Bessel series for $r$ (the
radius of the cylinder) and Fourier series for the $z$ (the thickness
of the zone). We then find that the positron flux is obtained as
\begin{eqnarray}
  \left[\Phi_{e^\pm}\right]_{\rm DM}
  =
  \frac{c}{4 \pi} f_{e^{\pm}}
  =
  \frac{c}{4\pi m_{\rm LSP} \tau_{\rm LSP}} 
  \int^\infty_E
  dE' G(E,E') \left[\frac{dN_{e^\pm}}{dE'} \right]_{\rm dec}.
\end{eqnarray}
Defining the the variable 
\begin{eqnarray}
  X(E,E') = \frac{E^{\delta - 1} - (E')^{\delta -  1}}{\delta - 1},
\end{eqnarray}
the Green's function $G(E,E')$ turns out to be
\begin{eqnarray}
  G(E,E')
  &=&
  \frac{\tau}{E^2} \sum_{n = 1}^\infty \sum_{m = 0}^\infty I_{nm}
  \exp\left[\left(\frac{\zeta_n^2}{R^2} + \frac{m^2\pi^2}{4L^2}\right)
    K_{e^\pm}^{(0)} \tau X(E,E')\right],
  \label{Green Fn}
  \\
  I_{nm}
  &=&
  \frac{2}{J_1^2(\zeta_n) R^2L}
  J_0\left[\frac{\zeta_n}{R} r_\odot\right] \sin\left[-\frac{m\pi}{2}\right]
  \nonumber \\
  &&
  \times
  \int^R_0 dr \int^L_0 dz \ 2r\rho_{\rm DM}(\sqrt{r^2 + z^2})
  J_0\left[\frac{\zeta_n}{R} r \right] \sin\left[\frac{m\pi}{2L}(z - L)\right],
  \nonumber \\
\end{eqnarray}
where $\tau \equiv E^2/b(E)$, $J_n$ is the n-th order Bessel function,
and $\zeta_n$ are successive zeros of $J_0$. The above formula is
useful to calculate the Green's function, except for the region $E
\simeq E'$. This is due to the exponential suppression in Eq.\
(\ref{Green Fn}). For the other energy region, we have checked that
the summations over $n$ and $m$ are well converged when we take the
Bessel (Fourier) series large enough.\footnote
{We have taken the summations up to $n, m = 100$ in our study.}

When $E \simeq E'$, the convergence becomes worse.  In fact, the
result is not converged for the cuspy profiles such as NFW and Moore
profiles even if we take the summations up to ${\cal O}(1000)$.
However, note that the flux for $E \simeq E'$ originates from the LSP
decay in the vicinity of the solar system.  Using the fact, we can
express the Green's function $G(E,E')$ in another form as
\begin{eqnarray}
  G(E,E')
  =
  \frac{\tau}{E^2} 
  \left[
    \exp\left(- K_{e^\pm}^{(0)} \tau X(E,E') \nabla^2\right) 
    \rho_{\rm DM}(r)
    \right]_{r=r_\odot}.
  \label{Green Fn2}
\end{eqnarray}
In contrast with the previous formula, this formula is valid in the
region $X(E,E') \ll 1$, namely $E \simeq E'$. By matching two formulas
in Eqs.\ (\ref{Green Fn}) and (\ref{Green Fn2}) at appropriate value
of $X(E,E')$, we can obtain the Green's function numerically without
wasting time for computation.

\end{document}